\documentclass[aps,preprint]{revtex4}%
\usepackage{amsmath}
\usepackage{amsfonts}
\usepackage{amssymb}
\usepackage{graphicx}%
\providecommand{\U}[1]{\protect\rule{.1in}{.1in}}

\begin{document}
\title{Number of particle creation and decoherence in the nonideal dynamical Casimir
effect at finite temperature}
\author{L. C. Celeri$^{\text{1}}$, F. Pascoal$^{\text{1}}$, M. A. de Ponte$^{\text{1}%
}$, and M. H. Y. Moussa$^{\text{2}}$}
\affiliation{$^{\text{1}}$ Departamento de F\'{\i}sica, Universidade Federal de S\~{a}o
Carlos, Via Washington Luis, km 235, S\~{a}o Carlos, 13565-905, SP, Brasil.}
\affiliation{$^{\text{2}}$ Instituto de F\'{\i}sica, Universidade de S\~{a}o Paulo, Caixa
Postal 369, 13566-590, S\~{a}o Carlos, SP, Brasil.}

\begin{abstract}
In this work we investigate the dynamical Casimir effect in a nonideal cavity
by deriving an effective Hamiltonian. We first compute a general expression
for the average number of particle creation, applicable for any law of motion
of the cavity boundary. We also compute a general expression for the linear
entropy of an arbitrary state prepared in a selected mode, also applicable for
any law of motion of the cavity boundary. As an application of our results we
have analyzed both the average number of particle creation and linear entropy
within a particular oscillatory motion of the cavity boundary. On the basis of
these expressions we develop a comprehensive analysis of the resonances in the
number of particle creation in the nonideal dynamical Casimir effect. We also
demonstrate the occurrence of resonances in the loss of purity of the initial
state and estimate the decoherence times associated with these resonances.

\end{abstract}

\maketitle

\section{Introduction}

\subsection{The dynamical Casimir effect}

The dynamical Casimir effect (DCE), by which particles are created and
annihilated due to accelerating boundaries that disturb the quantum vacuum,
has been extensively studied since the quantization by Moore \cite{Moore}, in
the early 1970s, of the radiation field in a cavity with moving, perfectly
reflecting boundaries.\textbf{\ }The problem of the electromagnetic field
quantization in a time-dependent dielectric medium inside the cavity was
solved two decades later by Dodonov and co-workers \cite{Dodonov}.
Interestingly enough, both the nonuniform motion of the boundaries \cite{4,5}
and the sudden change of the refractive index of the dielectric \cite{6,7}
produce similar effects, resulting in particle creation from the quantum
vacuum . The quantum statistical properties of the created photon, expected to
exhibit nonclassical features, has also been analyzed in Refs. \cite{8,9},
where a nonthermal distribution and squeezing were predicted.

A decade ago, Eberlein \cite{Eberlein}, following the reasoning by Schwinger
\cite{Schwinger}, conjectured,that the sonoluminescence phenomenon results
from particle creation due to moving boundaries between media of different
polarizability, attracting even more attention to the DCE. As pointed out by
P. Knight \cite{Knight}, the proof of Eberlein's conjecture would require a
demonstration that the created photon pairs emerging from the\ DCE have the
nonclassical statistics expected from a purely quantum effect. This
observation points up the importance of the quantum optical view of the DCE,
especially regarding the computation of the quantum statistical properties of
the created photons.

The thermal effects on the creation of particles under the influence of
time-dependent boundaries has also been investigated, either by taking into
account a formal model for the reservoir \cite{Soff1} or simply by assuming a
closed system to be initially at thermal equilibrium \cite{Soff2}. Regarding
the relevance of thermal effects for the experimental verification of the DCE,
it is demonstrated in Ref. \cite{Soff2} that finite temperatures can enhance
the number of particle creation\ by several orders of magnitude. In Ref.
\cite{Soff1}, the formal reservoir is modeled by assuming that one of the
cavity boundaries is a fixed leaky mirror while the other boundary moves, as
sketched in Fig. 1. The reservoir thus comprehends a discrete space of
eigenfrequencies generated by an additional boundary fixed far away from the
leaky mirror. Such a model for a lossy cavity was first adopted in Ref.
\cite{Scully}, envisaging, however, a static dissipative cavity. It is worth
noting that L. Parker \cite{Parker}, addressing the problem of particle
creation in expanding universes in the late 1960s, observed that the initial
presence of bosons tends to increase the number of bosons created by the
expansion mechanism --- similarly to the result reported in Ref. \cite{Soff2}
--- while the situation is reversed for fermions. The problem of the expanding
universe bears great similarity to the DCE and considerable efforts have been
devoted to this subject \cite{Carlitz,4,Davies,Fabio}.

The formulation, by Law \cite{Law}, of an effective Hamiltonian for the DCE,
enabling the dynamical description of the cavity field in the Schr\"{o}dinger
picture, also represents a significant contribution. Through this Hamiltonian,
which exhibits the essential features of the physical process, it becomes
possible to know the explicit form of the field state and to describe, in a
simplified form, the characteristic resonances which are also relevant to the
experimental verification of the DCE. Regarding a cavity with moving
boundaries, the quadratic structure of the effective Hamiltonian incorporates
the instantaneous modes of the cavity, parametrically amplified and coupled to
each other due to the moving boundaries. Generalizing Law$^{\text{'}}$s
procedure, in Ref. \cite{Soff1} the authors derive an effective Hamiltonian
for the DCE in a more realistic leaky (3+1)-dimensional cavity, the reservoir
being modeled as mentioned above.

In the present manuscript, reasoning by analogy with the derivation in Ref.
\cite{Soff1}, we obtain a dissipative counterpart of the effective
Hamiltonian\textbf{\ }introduced in Ref. \cite{Law}. From this Hamiltonian, we
compute a general expression for the average number of created particles which
--- differently from the expression derived in Ref. \cite{Soff1}, holding only
for the parametric resonance condition --- applies to any the law of motion
for the boundary. Together with the effective Hamiltonian, our expression for
the number of photon creation enables us to draw the whole scenario of the
emerging resonances. Moreover, we provide a comprehensive analysis of the
decoherence mechanisms within the nonideal DCE, along the lines discussed below.

\subsection{Nonideal DCE and decoherence}

More recently, the study of decoherence within the DCE has produced some
interesting results \cite{Dodonov1,Maia}, linking two topics that attract much
attention nowadays from theoretical \cite{Zurek,CL,QECC,PKM,Lidar,Mickel} and
experimental physics \cite{Haroche,Wineland}. As far as decoherence is
concerned, it is well established that any process inducing quantum
fluctuations in the evolution of a quantum system leads to the decoherence of
its superposition states \cite{Zurek,CL}. As a typical case, the inevitable
dissipative mechanisms accompanying the injection\ of noise from the reservoir
into the system drags its pure state into a statistical mixture. In this
particular situation, such noise injection comes entirely from the large
number of degrees of freedom modeling a multimode reservoir. When a few
degrees of freedom are coupled to the system of interest, Poincar\'{e}
recurrence takes place instead of the decoherence process \cite{Zurek}.

Focusing on a radiation mode inside a nonideal cavity with stationary mirrors,
it is the photon absorption by the mirrors (or the photon leakage in an open
cavity) that triggers the decoherence dynamics \cite{Walls,MMC}. The
technological search for higher-quality cavities is of no less interest than
the theoretical efforts to provide mechanisms to bypass decoherence in a
quantum information processor \cite{QECC,PKM,Lidar,Mickel}. Although the
protocols proposed to control or circumvent decoherence go far beyond the
requirements for conditions that weaken the system-reservoir coupling
\cite{Landauer,Unruh}, again, the practical efforts aimed at achieving quantum
information processing --- such as the miniaturization of physical ingredients
like laser beams and microcavities --- are no less challenging.

Similarly to photon absorption in a nonideal cavity, photon creation and
scattering at moving cavity boundaries is also a source of fluctuation
injection into the mode of interest. In fact, as mentioned above, all the
cavity modes are coupled together due to the moving boundaries \cite{Law} by
processes leading to photon creation and scattering. Consequently, all the
instantaneous cavity modes are subjected to injection of noise from the
remaining instantaneous modes acting as a reservoir. Therefore, in a situation
where a superposition state is prepared in a particular mode of a nonideal
cavity with moving boundaries, two distinct sources of decoherence take place:
the reservoir itself and the mechanism of amplification and multimode coupling
induced by the moving boundary. A detailed analysis of the decoherence process
under both sources of noise injection can reveal interesting features of both
the DCE and decoherence.

A particular case of decoherence of a superposition state prepared in a
selected mode of an ideal cavity with oscillating boundaries has already been
analyzed in Ref. \cite{Dodonov1}, in which the authors focus on the resonance
condition, where the oscillatory frequency of the boundaries is an integer
multiple of the fundamental mode eigenfrequency. Under this condition ---
which maximizes the photon creation number \cite{DKN} --- and in the absence
of a reservoir, the decoherence time of a superposition of coherent states
prepared in the fundamental mode is estimated by considering the coupling of
this mode only with its first-excited neighbor. In Ref. \cite{Maia}, it is
demonstrated that the DCE induces the decoherence of a superposition state of
a massive mirror in a harmonic potential, within a time scale which depends on
the energy of the state components, thus obeying the correspondence principle.

In the present study, similarly to our analysis of the average particle
creation number, we also approach the problem of decoherence from a general
scenario where a superposition state is prepared in a selected mode of a
nonideal cavity with a mirror undergoing an arbitrary motion. We take into
account the coupling of the selected instantaneous mode with all the remaining
cavity modes and not only with its nearest neighbors, even under the resonance
condition. As depicted below, our analysis is again based on the effective
dissipative Hamiltonian derived by analogy with the approaches in Refs.
\cite{Soff1} and \cite{Law}. We observe that the linear entropy associated
with a superposition state prepared in a selected mode exhibits resonances
similar to those which take place in the average number of particle creation.
Such resonances of the linear entropy reveal that the purity loss of the
prepared superposition occurs in appreciable rates only for specific values of
the detuning between the oscillatory frequency of the boundary and the
fundamental mode of the static cavity. The decoherence time of the prepared
state associated with these resonances is computed analytically.
Interestingly, out of the resonances, the purity loss or the decoherence
process can be disregarded within the nonideal DCE.

\section{An effective Hamiltonian for the nonideal DCE}

As discussed above, an effective Hamiltonian for the quantized field in a
cavity with a moving boundary was presented by Law \cite{Law}. This exhibits
the essential features of the DCE and enables the dynamic description of the
field modes in the Schr\"{o}dinger picture. Law assumed a dielectric medium in
the cavity, with a permittivity that varied in time and space, which we shall
disregard\textbf{.} However, following the reasoning in \cite{Soff1}, we also
consider a dissipative potential $V(x)=\gamma\delta(x)$ to model a dispersive
mirror of the cavity \cite{Scully}. Thus, our starting Lagrangian density for
a massless and neutral scalar radiation field $\phi(x,t)$ acted upon by a
dissipative force, is given by ($c=1$)%
\begin{equation}%
\mathcal{L}%
\left(  x,t\right)  =\frac{1}{2}\left\{  \left[  \dot{\phi}(x,t)\right]
^{2}-\left[  \partial_{x}\phi(x,t)\right]  ^{2}-\gamma\delta(x)\phi
^{2}(x,t)\right\}  \text{,} \label{1}%
\end{equation}
where the boundary conditions $\phi(-L_{0},t)=\phi\left[  q(t),t\right]  =0$,
with $q(t)$ representing an arbitrary law of motion of the mirror, must be
satisfied. Expanding the whole radiation field $\phi(x,t)$ --- across the
cavity plus the reservoir --- into a complete and orthonormal set of
instantaneous mode functions $\psi_{k}(x,t)$, we thus write $\phi
(x,t)=\sum_{k}Q_{k}(t)\psi_{k}(x,t)$. Since $\psi_{k}(x,t)$ must satisfy the
differential equation%
\[
\left[  \partial_{x,x}-\gamma\delta(x)\right]  \psi_{k}(x,t)=-\omega_{k}%
^{2}(t)\psi_{k}(x,t)\text{,}%
\]
under the conditions $\psi_{k}(-L_{0},t)=\psi_{k}\left[  q(t),t\right]  =0$,
the instantaneous eigenmodes $\psi_{k}(x,t)$ are given by%
\[
\psi_{k}(x,t)=\left\{
\begin{array}
[c]{c}%
\begin{array}
[c]{ccccccc}%
C_{k}\sin\left\{  \left[  (x-q\left(  t\right)  \right]  \omega_{k}\right\}  &
\text{for} & 0 & \leq & x & \leq & q\text{,}\\
R_{k}\sin\left[  (x+L_{0})\omega_{k}\right]  & \text{for} & -L_{0} & \leq &
x & \leq & 0\text{,}%
\end{array}
\\%
\begin{array}
[c]{cccccccccccc}
&  &  & 0 &  &  &  &  &  &  &  & \text{elsewhere,}%
\end{array}
\end{array}
\right.
\]
where the coefficients $C_{k}$ and $R_{k}$ are integration constants in $x$
defined by the normalization and the boundary conditions, while the
eigenfrequencies $\omega_{k}(q)$ are computed from the transcendental equation%
\begin{equation}
\cot\left[  \omega_{k}q\left(  t\right)  \right]  +\cot\left(  \omega_{k}%
L_{0}\right)  =-\gamma/\omega_{k}\text{,} \label{2}%
\end{equation}
which results from the continuity conditions on the static mirror at $x=0$.
Following Ref. \cite{Soff1}, it is possible to obtain an approximate
analytical solution for Eq. (\ref{2}) in the limit where the lossy mirror is
nearly ideal, yielding the perturbative parameter $\eta_{k}=\omega_{k}%
/\gamma\ll1$. Under the assumption that the ratio $q\left(  t\right)  /L_{0}$
is a noninteger number --- which is true for the situation in hand, where
$L_{0}\gg q\left(  t\right)  $ --- we find that only one of the cotangent
functions in (\ref{2}) becomes dominating. Therefore, by expanding one of
these functions around its poles $n\pi$, $n$ being an integer, we obtain a
polynomial that can be solved for $\omega_{k}$ as a series expansion in
$\eta_{k}$. Depending on which is the dominating cotangent function, we obtain
two classes of eigenfunctions $\psi_{k}(x,t)$, respectively derived from the
cavity-dominated and the reservoir-dominated eigenfrequencies, which, to first
order in $\eta_{k}$, reduce to
\begin{subequations}
\label{3}%
\begin{align}
\omega_{k}^{\mathcal{C}}(q)  &  =\frac{k\pi}{q}\left(  1+\frac{1}{\gamma
q}\right)  ^{-1}\text{,}\label{3a}\\
\omega_{k}^{\mathcal{R}}(L_{0})  &  =\frac{k\pi}{L_{0}}\left(  1+\frac
{1}{\gamma L_{0}}\right)  ^{-1}\text{.} \label{3b}%
\end{align}
We note that, under the above approximation, the reservoir eigenfrequency
$\omega_{k}^{\mathcal{R}}$ does not depend on time. Evidently, these
eigenfrequencies define cavity-dominated and reservoir-dominated eigenmodes,
$\psi_{k}^{\mathcal{C}}(x,t)$ and $\psi_{k}^{\mathcal{R}}(x,t)$.

Substituting the expanded $\phi(x,t)$ into Eq. (\ref{1}) and integrating the
result over all space --- between $x=-L_{0}$\ and $x=q(t)$ --- we obtain the
following Lagrangian function
\end{subequations}
\[
\mathcal{L}=\frac{1}{2}\sum_{k}\left[  \dot{Q}_{k}^{2}-\omega_{k}^{2}%
(t)Q_{k}^{2}+Q_{k}\sum_{\ell}G_{k\ell}(t)\left(  \dot{Q}_{\ell}+\sum_{m}%
Q_{m}G_{m\ell}(t)\right)  \right]  \text{,}%
\]
where the antisymmetric coefficients $G_{k\ell}$ are given by%
\[
G_{k\ell}(t)=-G_{\ell k}(t)=\int_{-L_{0}}^{q(t)}dx\text{ }\dot{\psi}%
_{k}(x,t)\psi_{\ell}(x,t)\text{.}%
\]
Introducing the canonical conjugated momenta $P_{k}=\partial L/\partial\dot
{Q}_{k}$, we obtain from the Legendre transformation, the Hamiltonian
\begin{equation}
\mathcal{H}=\frac{1}{2}\sum_{k}\left[  P_{k}^{2}+\omega_{k}^{2}(t)Q_{k}%
^{2}+2\sum_{\ell}P_{k}G_{k\ell}(t)Q_{\ell}\right]  \text{,} \label{4}%
\end{equation}
which gives the coupled equation of motion
\begin{subequations}
\label{5}%
\begin{align}
\dot{Q}_{k}  &  =P_{k}+%
{\displaystyle\sum\limits_{\ell}}
G_{k\ell}Q_{\ell}\text{,}\label{5a}\\
\dot{P}_{k}  &  =-\omega_{k}^{2}Q_{k}+%
{\displaystyle\sum\limits_{\ell}}
G_{k\ell}P_{\ell}\text{.} \label{5b}%
\end{align}

\subsection{Instantaneous photon creation and annihilation operators}

In order to study the phenomenon of photon creation --- through the quantum
version of the above Hamiltonian (\ref{4}) --- it is convenient to introduce,
as in Ref. \cite{Law}, the \textquotedblleft instantaneous\textquotedblright%
\ annihilation and creation operators
\end{subequations}
\begin{align*}
a_{k}(t)  &  =\frac{1}{\sqrt{2\omega_{k}(t)}}\left[  \omega_{k}(t)Q_{k}%
+iP_{k}\right]  \text{,}\\
a_{k}^{\dagger}(t)  &  =\frac{1}{\sqrt{2\omega_{k}(t)}}\left[  \omega
_{k}(t)Q_{k}-iP_{k}\right]  \text{,}%
\end{align*}
which act on the radiation field in the whole space, including the cavity and
the reservoir, and satisfy the equal-time commutation relation $\left[
a_{k}(t),a_{\ell}^{\dagger}(t)\right]  =\delta_{k\ell}$. The time derivative
of these ladder operators, combined with Eqs. (\ref{5}), give us $\dot{a}_{k}$
and $\dot{a}_{k}^{\dagger}$ as linear functions of $Q_{k}$ and $P_{k}$ and,
consequently, of $a_{k}$ and $a_{k}^{\dagger}$. Through the Heisenberg
equation of motion we thus infer a general quadratic form for the effective
Hamiltonian that governs the motion of $a_{k}$ and $a_{k}^{\dagger}$, which
turns out to be that derived by Law \cite{Law}
\begin{equation}
H_{eff}=\sum_{k}\left\{  \omega_{k}(t)a_{k}^{\dagger}a_{k}+i\xi_{k}(t)\left[
(a_{k}^{\dagger})^{2}-a_{k}^{2}\right]  +i\sum_{\ell(\neq k)}\mu_{k\ell
}(t)\left(  a_{k}^{\dagger}a_{\ell}^{\dagger}+a_{k}^{\dagger}a_{\ell
}-H.c.\right)  \right\}  \text{,} \label{6}%
\end{equation}
but with the dissipative mechanism being introduced through the time-dependent
frequency $\omega_{k}(t)$. The dissipation is also taken into account in the
strengths $\xi_{k}(t)$ and $\mu_{k\ell}(t)$ --- associated with degenerate
$\left[  (a_{k}^{\dagger})^{2}-a_{k}^{2}\right]  $ and nondegenerate $\left(
a_{k}^{\dagger}a_{\ell}^{\dagger}-a_{k}a_{\ell}\right)  $ photon-pair
creation, apart from photon scattering $\left(  a_{k}^{\dagger}a_{\ell}%
-a_{k}a_{\ell}^{\dagger}\right)  $ --- given by
\begin{subequations}
\label{7}%
\begin{align}
\xi_{k}(t)  &  =\frac{\dot{\omega}_{k}(t)}{4\omega_{k}(t)}\text{,}\label{7a}\\
\mu_{k\ell}(t)  &  =\frac{1}{2}\sqrt{\frac{\omega_{k}(t)}{\omega_{\ell}(t)}%
}G_{k\ell}(t)\text{.} \label{7b}%
\end{align}

Through the unitary transformation $U(t)=\exp\left(  -i%
{\textstyle\int\nolimits_{0}^{t}}
d\tau\text{ }H_{0}(\tau)\right)  $, where $H_{0}=%
{\textstyle\sum\nolimits_{k}}
\omega_{k}(t)a_{k}^{\dagger}a_{k}$ such that $\left[  U(t),H_{0}(t)\right]
=0$, we rewrite Hamiltonian (\ref{6}) in the interaction picture%
\end{subequations}
\[
H_{I}=i\sum_{k}\left\{  \operatorname*{e}\nolimits^{i\Omega_{k}(t)}\left[
\xi_{k}(t)\left(  a_{k}^{\dagger}\right)  ^{2}\operatorname*{e}%
\nolimits^{i\Omega_{k}(t)}+\sum_{\ell(\neq k)}\mu_{k\ell}(t)a_{k}^{\dagger
}\left(  a_{\ell}^{\dagger}\operatorname*{e}\nolimits^{i\Omega_{\ell}%
(t)}+a_{\ell}\operatorname*{e}\nolimits^{-i\Omega_{\ell}(t)}\right)  \right]
-H.c.\right\}  \text{,}%
\]
where $\Omega_{k}(t)=\int_{0}^{t}d\tau$ $\omega_{k}(\tau)$. Defining the
operator $\mathcal{O}_{\mathcal{S},k}(t)=a_{\mathcal{S},k}\exp\left[
-i\Omega_{k}^{\mathcal{S}}(t)\right]  $, the degenerate term of the above
Hamiltonian can be decomposed into two components, for the cavity
($\mathcal{S}=\mathcal{C}$)- and the reservoir ($\mathcal{S}=\mathcal{R}%
$)-dominated modes, given $\xi_{k}^{\mathcal{C}}(t)\left[  \left(
\mathcal{O}_{\mathcal{C},k}^{\dagger}(t)\right)  ^{2}-H.c.\right]  +\xi
_{k}^{\mathcal{R}}(t)\left[  \left(  \mathcal{O}_{\mathcal{R},k}^{\dagger
}(t)\right)  ^{2}-H.c.\right]  $, where $\xi_{k}^{\mathcal{S}}(t)=\dot{\omega
}_{k}^{\mathcal{S}}(t)/4\omega_{k}^{\mathcal{S}}(t)$. However, to first order
in $\eta_{k}$ the reservoir eigenfrequency $\omega_{k}^{\mathcal{R}}(t)$ does
not depend on time, so that from Eq. (\ref{7a}) we obtain $\xi_{k}%
^{\mathcal{R}}(t)=0$, implying that there is no contribution of the degenerate
term in a reservoir-dominated mode. As far as the nondegenerate and scattering
terms are concerned, they can be decomposed into four components of the form
$\mu_{k\ell}^{\mathcal{SS}^{\prime}}(t)\left[  \mathcal{O}_{\mathcal{S}%
,k}^{\dagger}(t)\mathcal{O}_{\mathcal{S}^{\prime},\ell}^{\dagger
}(t)+\mathcal{O}_{\mathcal{S},k}^{\dagger}(t)\mathcal{O}_{\mathcal{S}^{\prime
},\ell}(t)-H.c.\right]  $, with $\mathcal{S}$,$\mathcal{S}^{\prime
}=\mathcal{C}$ or $\mathcal{R}$. The components $\mathcal{S}=\mathcal{S}%
^{\prime}=\mathcal{C}$ ($\mathcal{R}$) account for the interaction between two
cavity (reservoir)-dominated modes, whereas the components $\mathcal{S}%
\neq\mathcal{S}^{\prime}$ implies interaction between a cavity- and a
reservoir-dominated mode. The double-labeled strengths are given by
$\mu_{k\ell}^{\mathcal{SS}^{\prime}}(t)=\sqrt{\omega_{k}^{\mathcal{S}%
}(t)/\omega_{\ell}^{\mathcal{S}^{\prime}}(t)}G_{k\ell}^{\mathcal{SS}^{\prime}%
}(t)/2$ where $G_{k\ell}^{\mathcal{SS}^{\prime}}(t)=\int_{-L_{0}}^{q(t)}dx$
$\dot{\psi}_{k}^{\mathcal{S}}(x,t)\psi_{\ell}^{\mathcal{S}^{\prime}}(x,t)$.
Evidently, since to first order in $\eta_{k}$ neither the reservoir-dominated
mode $\psi_{k}^{\mathcal{R}}(x,t)$ nor $\omega_{k}^{\mathcal{R}}(t)$ depends
on time, it follows that $G_{k\ell}^{\mathcal{RR}}(t)=G_{k\ell}^{\mathcal{RC}%
}(t)=0$ and, consequently, $\mu_{k\ell}^{\mathcal{RR}}(t)=\mu_{k\ell
}^{\mathcal{RC}}(t)=0$. Therefore, we arrive at the final effective
Hamiltonian%
\begin{align}
H_{I}  &  =i\sum_{k}\left\{  \xi_{k}^{\mathcal{C}}(t)\left(  \mathcal{O}%
_{\mathcal{C},k}^{\dagger}(t)\right)  ^{2}+\sum_{\ell(\neq k)}\left\{
\mu_{k\ell}^{\mathcal{CC}}(t)\mathcal{O}_{\mathcal{C},k}^{\dagger}(t)\left[
\mathcal{O}_{\mathcal{C},\ell}^{\dagger}(t)+\mathcal{O}_{\mathcal{C},\ell
}(t)\right]  \right.  \right. \nonumber\\
&  \left.  \left.  +\mu_{k\ell}^{\mathcal{CR}}(t)\mathcal{O}_{\mathcal{C}%
,k}^{\dagger}(t)\left[  \mathcal{O}_{\mathcal{R},\ell}^{\dagger}%
(t)+\mathcal{O}_{\mathcal{R},\ell}(t)\right]  \right\}  -H.c.\right\}
\text{,} \label{8}%
\end{align}
which, apart from accounting for dissipation, also provides, as does
Law$^{\text{'}}$s effective Hamiltonian, the expected weak quadratic
amplification of the cavity leaking modes and the weak coupling between all
these dynamical modes. In fact, from Eq. (\ref{7}) we find that both strengths
$\xi_{k}(t)$ and $\mu_{k\ell}(t)$ --- proportional to the velocity of the
boundary $\dot{q}(t)$ which is small in nonrelativistic cases --- are
significantly smaller than the fundamental modes of the static cavity. We also
note that the above-derived dissipative effective Hamiltonian, Eq. (\ref{8}),
generalizes that derived in Ref. \cite{Soff1} in that it accounts for any law
of motion for the boundary. In Ref. \cite{Soff1}, the authors consider a
specific oscillatory law of motion and proceed to the analysis of photon
creation only within resonance conditions. In the present approach, following
the development presented below, we obtain an expression for the average
number of photon creation for any law of motion of the boundary whatsoever,
which enables us to present a comprehensive analysis of the resonances on the
number of photon creation in the nonideal DCE.

\section{The average number of photon creation}

Having derived the effective Hamiltonian (\ref{8}), the evolution of the
density operator of the system $\rho(t)$, in the interaction picture, is
determined by the equation $\dot{\rho}(t)=-i\left[  H_{I},\rho(t)\right]  $.
Taking the velocity $\dot{q}(t)$ as our perturbative parameter, the formal
solution of $\rho(t)$ coming from terms up to second order in $H_{I}(t)$,
reads%
\[
\rho(t)\simeq\rho(0)-i\int_{0}^{t}d\tau\text{ }\left[  H_{I}(\tau
),\rho(0)\right]  -\int_{0}^{t}dt^{\prime}\int_{0}^{t^{\prime}}d\tau\text{
}\left[  H_{I}(t^{\prime}),\left[  H_{I}(\tau),\rho(0)\right]  \right]
\text{.}%
\]

We assume all reservoir-dominated modes to be in a thermal state
$\rho_{\mathcal{R}}(0)=e^{-\beta H_{\mathcal{R}}}/\operatorname{Tr}\left(
e^{-\beta H_{\mathcal{R}}}\right)  $, where $H_{\mathcal{R}}=%
{\textstyle\sum\nolimits_{k}}
\omega_{k}^{\mathcal{R}}(t)a_{\mathcal{R},k}^{\dagger}a_{\mathcal{R},k}$,
$\beta=1/k_{B}T$, $k_{B}$ being the Boltzmann constant and $T$ the temperature
of the reservoir. For the cavity-dominated modes we consider any initial state
$\rho_{\mathcal{C}}(0)=%
{\textstyle\prod\nolimits_{k}}
\rho_{\mathcal{C},k}(0)$ under the conditions that all these modes but the
$k$th satisfy the relations $\operatorname{Tr}_{\mathcal{C}\left(  \neq
k\right)  }\left[  \left(  a_{\mathcal{C},\ell}\right)  ^{m}\left(
a_{\mathcal{C},\ell^{\prime}}^{\dagger}\right)  ^{n}\rho_{\mathcal{C}%
}(0)\right]  $,$\operatorname{Tr}_{\mathcal{C}\left(  \neq k\right)  }\left[
\left(  a_{\mathcal{C},\ell}^{\dagger}\right)  ^{m}\left(  a_{\mathcal{C}%
,\ell^{\prime}}\right)  ^{n}\rho_{\mathcal{C}}(0)\right]  \propto\delta
_{mn}\delta_{\ell\ell^{\prime}}$ and $\operatorname{Tr}_{\mathcal{C}\left(
\neq k\right)  }\left[  \left(  a_{\mathcal{C},\ell}\right)  ^{m}\left(
a_{\mathcal{C},\ell^{\prime}}\right)  ^{n}\rho_{\mathcal{C}}(0)\right]
$,$\operatorname{Tr}_{\mathcal{C}\left(  \neq k\right)  }\left[  \left(
a_{\mathcal{C},\ell}^{\dagger}\right)  ^{m}\left(  a_{\mathcal{C},\ell
^{\prime}}^{\dagger}\right)  ^{n}\rho_{\mathcal{C}}(0)\right]  =0$. In
particular, we may assume the realistic condition that all the
cavity-dominated modes but the $k$th, are in thermal states $\rho
_{\mathcal{C},\ell}(0)=e^{-\beta H_{\mathcal{C},\ell}}/\operatorname{Tr}%
\left(  e^{-\beta H_{\mathcal{C},\ell}}\right)  $, with $H_{\mathcal{C},\ell
}=\omega_{\ell}^{\mathcal{C}}(t)a_{\mathcal{C},\ell}^{\dagger}a_{\mathcal{C}%
,\ell}$, as like as the reservoir-dominated modes. For the $k$th mode --- the
one for which we compute the average number of photon creation --- we consider
any arbitrary initial state $\rho_{\mathcal{C},k}(0)$, including the thermal distribution.

We next compute the average number of photon creation in a particular
cavity-dominated mode $k$, given by $\left\langle N_{\mathcal{C}%
,k}(t)\right\rangle =\operatorname{Tr}\left[  \rho_{\mathcal{C},k}%
(t)a_{\mathcal{C},k}^{\dagger}a_{\mathcal{C},k}\right]  $, the trace being
taken over the $k$th cavity-dominated mode. Since the interaction Hamiltonian
$H_{I}$ is null for a static cavity, all the cavity- and reservoir-dominated
modes are initially uncorrelated, such that $\rho(0)=\rho_{\mathcal{C}%
}(0)\otimes\rho_{\mathcal{R}}(0)$. By computing the reduced operator%
\[
\rho_{\mathcal{C},k}(t)=\operatorname{Tr}_{\mathcal{R}}\operatorname{Tr}%
_{\mathcal{C}\left(  \neq k\right)  }\rho(t)=\sum_{\left\{  n_{\mathcal{R}%
,k}\right\}  }\sum_{\left\{  n_{\mathcal{C},\ell\left(  \neq k\right)
}\right\}  }\left\langle \left\{  n_{\mathcal{R},k}\right\}  \right\vert
\left\langle \left\{  n_{\mathcal{C},\ell}\right\}  \left\vert \rho
(t)\right\vert \left\{  n_{\mathcal{C},\ell}\right\}  \right\rangle \left\vert
\left\{  n_{\mathcal{R},k}\right\}  \right\rangle \text{,}%
\]
where the trace is taken over all the reservoir- and cavity-dominated modes,
except the $k$th cavity-dominated mode, we obtain%
\begin{align}
\rho_{\mathcal{C},k}(t)  &  =\rho_{\mathcal{C},k}(0)-i\int_{0}^{t}dt^{\prime
}\text{ }\left[  V_{\mathcal{C},k}(t^{\prime}),\rho_{\mathcal{C},k}(0)\right]
\nonumber\\
&  -\int_{0}^{t}dt^{\prime}\int_{0}^{t^{\prime}}d\tau\text{ }\left\{  \left[
V_{\mathcal{C},k}(t^{\prime}),\left[  V_{\mathcal{C},k}(\tau),\rho
_{\mathcal{C},k}(0)\right]  \right]  \right. \nonumber\\
&  -\text{ }\sum_{\ell(\neq k)}\left\{  \left[  f_{\mathcal{C},\ell}%
(t^{\prime},\tau)\Xi_{\ell k}^{+}(t^{\prime},\tau)+g_{\ell}(t^{\prime}%
,\tau)\Upsilon_{\ell k}^{+}(t^{\prime},\tau)\right]  \left[  \mathcal{O}%
_{\mathcal{C},k}(t^{\prime}),\mathcal{O}_{\mathcal{C},k}^{\dagger}(\tau
)\rho_{\mathcal{C},k}(0)\right]  \right. \nonumber\\
&  +\left[  f_{\mathcal{C},\ell}(t^{\prime},\tau)\Xi_{\ell k}^{+}(t^{\prime
},\tau)-g_{\ell}(t^{\prime},\tau)\Upsilon_{\ell k}^{+}(t^{\prime}%
,\tau)\right]  \left[  \mathcal{O}_{\mathcal{C},k}^{\dagger}(t^{\prime
}),\mathcal{O}_{\mathcal{C},k}(\tau)\rho_{\mathcal{C},k}(0)\right] \nonumber\\
&  +\left[  f_{\mathcal{C},\ell}(t^{\prime},\tau)\Xi_{\ell k}^{-}(t^{\prime
},\tau)+g_{\ell}(t^{\prime},\tau)\Upsilon_{\ell k}^{-}(t^{\prime}%
,\tau)\right]  \left[  \mathcal{O}_{\mathcal{C},k}^{\dagger}(t^{\prime
}),\mathcal{O}_{\mathcal{C},k}^{\dagger}(\tau)\rho_{\mathcal{C},k}(0)\right]
\nonumber\\
&  \left.  +\left[  f_{\mathcal{C},\ell}(t^{\prime},\tau)\Xi_{\ell k}%
^{-}(t^{\prime},\tau)-g_{\ell}(t^{\prime},\tau)\Upsilon_{\ell k}^{-}%
(t^{\prime},\tau)\right]  \left[  \mathcal{O}_{\mathcal{C},k}(t^{\prime
}),\mathcal{O}_{\mathcal{C},k}(\tau)\rho_{\mathcal{C},k}(0)\right]
+H.c.\right\} \nonumber\\
&  \left.  -\sum_{\ell}\mu_{k\ell}^{C\mathcal{R}}(t^{\prime})\mu_{k\ell
}^{C\mathcal{R}}(\tau)\left\{  f_{\mathcal{R},\ell}(t^{\prime},\tau)\left[
\Lambda_{\mathcal{C},k}(t^{\prime}),\Lambda_{\mathcal{C},k}(\tau
)\rho_{\mathcal{C},k}(0)\right]  +H.c.\right\}  \right\}  \text{,} \label{8.5}%
\end{align}
where we have defined the time-dependent operators%
\begin{align*}
V_{\mathcal{C},k}(t)  &  =i\xi_{k}^{\mathcal{C}}(t)\left[  \left(
\mathcal{O}_{\mathcal{C},k}^{\dagger}(t)\right)  ^{2}-\mathcal{O}%
_{\mathcal{C},k}^{2}(t)\right]  \text{,}\\
\Lambda_{\mathcal{C},k}(t)  &  =\mathcal{O}_{\mathcal{C},k}^{\dagger
}(t)-\mathcal{O}_{\mathcal{C},k}(t)\text{,}%
\end{align*}
and functions%
\begin{align*}
f_{\mathcal{S},\ell}(t,\tau)  &  =-2N_{\mathcal{S},\ell}(0)\cos\Delta_{\ell
}^{\mathcal{S}}(t,\tau)-\exp\left[  -i\Delta_{\ell}^{\mathcal{S}}%
(t,\tau)\right]  \text{,}\\
g_{\ell}(t,\tau)  &  =2iN_{\mathcal{C},\ell}(0)\sin\Delta_{\ell}^{\mathcal{C}%
}(t,\tau)-\exp\left[  -i\Delta_{\ell}^{\mathcal{C}}(t,\tau)\right]  \text{,}\\
\Xi_{\ell k}^{\pm}(t,\tau)  &  =\zeta_{\ell k}(t)\zeta_{\ell k}(\tau)\pm
\zeta_{k\ell}(t)\zeta_{k\ell}(\tau)\text{,}\\
\Upsilon_{\ell k}^{\pm}(t,\tau)  &  =\zeta_{\ell k}(t)\zeta_{k\ell}(\tau
)\pm\zeta_{k\ell}(t)\zeta_{\ell k}(\tau)\text{,}\\
\Delta_{\ell}^{\mathcal{S}}(t,\tau)  &  =\Omega_{\ell}^{\mathcal{S}}%
(t)-\Omega_{\ell}^{\mathcal{S}}(\tau)\text{,}\\
\zeta_{\ell k}(t)  &  =\xi_{k}^{\mathcal{C}}(t)\delta_{\ell k}+\mu_{k\ell
}^{\mathcal{CC}}(t)\text{,}%
\end{align*}
in which $N_{\mathcal{S},\ell}(0)=\operatorname{Tr}\left(  \rho_{\mathcal{S}%
,\ell}(0)a_{\mathcal{S},\ell}^{\dagger}a_{\mathcal{S},\ell}\right)  $
represents the initial average number of photons in the $\mathcal{S}%
$-dominated mode $\ell$. (Particularly, when all the cavity-dominated modes
are initially in thermal states, we get $N_{\mathcal{S},\ell}(0)=1/\left[
\exp\left(  \beta\omega_{\ell}^{\mathcal{S}}\right)  -1\right]  $.)

From the above expression for the reduced operator $\rho_{\mathcal{C},k}(t)$,
we finally obtain the average number of photon creation in the
cavity-dominated $k$ mode $\Delta N_{\mathcal{C},k}(t)=\left\langle
N_{\mathcal{C},k}(t)\right\rangle -N_{\mathcal{C},k}(0)$, given by
\begin{equation}
\Delta N_{\mathcal{C},k}(t)=2\operatorname{Re}\int_{0}^{t}dt^{\prime}\int
_{0}^{t^{\prime}}d\tau\text{ }\left\{  \mathcal{F}_{k}(t^{\prime}%
,\tau)+\mathcal{G}_{k}(t^{\prime},\tau)\right\}  \text{,} \label{9}%
\end{equation}
where we have defined the time-dependent functions
\begin{align*}
\mathcal{F}_{k}(t,\tau)  &  =\sum_{\ell}\left\{  N_{\mathcal{C},k}(0)\left[
f_{\mathcal{C},\ell}(t,\tau)\Xi_{\ell k}^{+}(t,\tau)-g_{\ell}(t,\tau
)\Upsilon_{\ell k}^{+}(t,\tau)\right]  \exp\left[  -i\Delta_{k}^{\mathcal{C}%
}(t,\tau)\right]  \right. \\
&  \left.  -\left[  N_{\mathcal{C},k}(0)+1\right]  \left[  f_{\mathcal{C}%
,\ell}(t,\tau)\Xi_{\ell k}^{+}(t,\tau)+g_{\ell}(t,\tau)\Upsilon_{\ell k}%
^{+}(t,\tau)\right]  \exp\left[  i\Delta_{k}^{\mathcal{C}}(t,\tau)\right]
\right\}  \text{,}\\
\mathcal{G}_{k}(t,\tau)  &  =\sum_{\ell}\mu_{k\ell}^{\mathcal{CR}}%
(t)\mu_{k\ell}^{\mathcal{CR}}(\tau)\left\{  \left[  2N_{\mathcal{C}%
,k}(0)+1\right]  \sin\left[  \Delta_{\ell}^{\mathcal{C}}(t,\tau)\right]
\sin\left[  \Delta_{\ell}^{\mathcal{R}}(t,\tau)\right]  \right. \\
&  \left.  -\left[  2N_{\mathcal{C},\ell}(0)+1\right]  \cos\left[
\Delta_{\ell}^{\mathcal{C}}(t,\tau)\right]  \cos\left[  \Delta_{\ell
}^{\mathcal{R}}(t,\tau)\right]  \right\}  \text{.}%
\end{align*}

For the particular case of an ideal cavity ($\gamma\rightarrow\infty$) at
absolute zero ($N_{\mathcal{S},k}(T=0)\rightarrow0$), the above expression for
the average number of photon creation simplifies to%
\[
\Delta N_{\mathcal{C},k}(t)=2\int_{0}^{t}dt^{\prime}\int_{0}^{t^{\prime}}%
d\tau\text{ }\sum_{\ell}\chi_{k\ell}(t^{\prime},\tau)\cos\left[  \Omega
_{k}^{\mathcal{C}}(t^{\prime})-\Omega_{k}^{\mathcal{C}}(\tau)+\Omega_{\ell
}^{\mathcal{C}}(t^{\prime})-\Omega_{\ell}^{\mathcal{C}}(\tau)\right]  \text{,}%
\]
with%
\[
\chi_{k\ell}\left(  t,\tau\right)  =\frac{\dot{q}\left(  t\right)  \dot
{q}\left(  \tau\right)  }{q\left(  t\right)  q\left(  \tau\right)  }%
\times\left\{
\begin{array}
[c]{ccc}%
1/8 & \text{for} & k=\ell\text{,}\\
\frac{k\ell}{\left(  k+\ell\right)  ^{2}} & \text{for} & k\neq\ell\text{.}%
\end{array}
\right.
\]
Finally, for a static cavity ($\dot{q}=0$), it is straightforward to verify
that there is no photon creation, so that $\left\langle N_{\mathcal{C}%
,k}(t)\right\rangle =N_{\mathcal{C},k}(0)$.

\section{Linear entropy and decoherence}

To analyze the decoherence of quantum states in the dynamical Casimir effect,
suppose that an arbitrary superposition state $\rho_{\mathcal{C},k}(0)$ is
prepared in the $k$th cavity-dominated mode of an initially static cavity.
Evidently, due to the motion of the cavity boundary --- which induces the
coupling of the$\ $selected mode with all other modes of the cavity apart from
the reservoir --- the linear entropy of the evolved state $S_{\mathcal{C}%
,k}(t)=1-\operatorname*{Tr}\rho_{\mathcal{C},k}^{2}(t)$ must increase. We
stress that this behavior, associated with the purity loss of the initial
state $\rho_{\mathcal{C},k}(0)$, occurs even when scattering is the only
coupling mechanism arising from the motion of the mirror. However, when the
photon creation process also takes place, the purity loss is significantly
increased, as we demonstrate below.

To compute the linear entropy $S_{\mathcal{C},k}(t)$ we confine ourselves to
the case of an absolute zero \textquotedblleft cavity $+$
reservoir\textquotedblright\ system, where all the reservoir- and
cavity-dominated modes, except the $k$th cavity-dominated mode, are in the
vacuum state, i.e,%
\begin{equation}
\rho(0)=\left\vert \left\{  0_{\mathcal{C},\ell\left(  \neq k\right)
}\right\}  \right\rangle \left\langle \left\{  0_{\mathcal{C},\ell\left(  \neq
k\right)  }\right\}  \right\vert \otimes\left\vert \left\{  0_{\mathcal{R}%
,\ell}\right\}  \right\rangle \left\langle \left\{  0_{\mathcal{R},\ell
}\right\}  \right\vert \otimes\rho_{\mathcal{C},k}(0). \label{10}%
\end{equation}
Therefore, from the reduced density operator (\ref{8.5}), with $\rho(0)$ given
by (\ref{10}), we obtain, up to second-order in $\dot{q}$, the result%
\begin{align}
S_{\mathcal{C},k}(t)  &  =1-\operatorname*{Tr}\rho_{\mathcal{C},k}%
^{2}(0)\nonumber\\
&  +4\operatorname{Re}\int_{0}^{t}dt^{\prime}\int_{0}^{t^{\prime}}%
d\tau\left\{  \sum_{\ell(\neq k)}\left[  \Xi_{\ell k}^{+}(t^{\prime}%
,\tau)\left\langle \Theta_{k}^{+}(t^{\prime},\tau)\right\rangle -\Upsilon
_{\ell k}^{+}(t^{\prime},\tau)\left\langle \Theta_{k}^{-}(t^{\prime}%
,\tau)\right\rangle \right.  \right. \nonumber\\
&  \left.  +\Xi_{\ell k}^{-}(t^{\prime},\tau)\left\langle \Phi_{k}%
^{+}(t^{\prime},\tau)\right\rangle +\Upsilon_{\ell k}^{-}(t^{\prime}%
,\tau)\left\langle \Phi_{k}^{-}(t^{\prime},\tau)\right\rangle \right]
\exp\left[  -i\Delta_{\ell}^{\mathcal{C}}(t^{\prime},\tau)\right] \nonumber\\
&  \left.  +\sum_{\ell}\mu_{k\ell}^{C\mathcal{R}}(t^{\prime})\mu_{k\ell
}^{C\mathcal{R}}(\tau)\left[  \left\langle \Theta_{k}^{+}(t^{\prime}%
,\tau)\right\rangle -\left\langle \Phi_{k}^{+}(t^{\prime},\tau)\right\rangle
\right]  \exp\left[  -i\Delta_{\ell}^{\mathcal{R}}(t^{\prime},\tau)\right]
\right\}  \text{,} \label{11}%
\end{align}
where we have defined the functions%
\begin{align*}
\left\langle \Theta_{k}^{\pm}(t,\tau)\right\rangle  &  =\operatorname*{Tr}%
\left\{  \left(  \left[  \mathcal{O}_{\mathcal{C},k}^{\dagger}(t),\mathcal{O}%
_{\mathcal{C},k}(\tau)\rho_{\mathcal{C},k}(0)\right]  \pm\left[
\mathcal{O}_{\mathcal{C},k}(t),\mathcal{O}_{\mathcal{C},k}^{\dagger}(\tau
)\rho_{\mathcal{C},k}(0)\right]  \right)  \rho_{\mathcal{C},k}(0)\right\}
\text{,}\\
\left\langle \Phi_{k}^{\pm}(t,\tau)\right\rangle  &  =\operatorname*{Tr}%
\left\{  \left(  \left[  \mathcal{O}_{\mathcal{C},k}^{\dagger}(t),\mathcal{O}%
_{\mathcal{C},k}^{\dagger}(\tau)\rho_{\mathcal{C},k}(0)\right]  \pm\left[
\mathcal{O}_{\mathcal{C},k}(t),\mathcal{O}_{\mathcal{C},k}(\tau)\rho
_{\mathcal{C},k}(0)\right]  \right)  \rho_{\mathcal{C},k}(0)\right\}  \text{.}%
\end{align*}
Evidently, for a static cavity, $\dot{q}=0$, we end up with $S_{\mathcal{C}%
,k}(t)=1-\operatorname*{Tr}\rho_{\mathcal{C},k}^{2}(0)$. We finally stress
that the decoherence time $\tau_{D}$ of an initial state $\rho_{\mathcal{C}%
,k}(0)$ can be estimated from the entropy in Eq. (\ref{11}), as discussed below.

\section{Phenomenology of the effective Hamiltonian for an oscillatory law of
motion of the boundary}

Before analyzing the mechanisms of the photon creation and linear entropy,
defined in expressions (\ref{9}) and (\ref{11}), it is worth considering the
phenomenological implications of the effective Hamiltonian (\ref{8}). This
will allow us to map out the main features arising from the general formulas
(\ref{9}) and (\ref{11}). To this end, let us specify a particular law of
motion for the mirror, namely the sinoidal law which maximizes the number of
photons created \cite{DKN}, given by%
\begin{equation}
q(t)=q_{0}\left[  1+\varepsilon\sin\left(  \mathfrak{p}\omega_{1}t\right)
\right]  \text{,} \label{12}%
\end{equation}
where $\varepsilon\ll1$ and $\left\vert \mathfrak{p}-1\right\vert \omega_{1}$
is the detuning between the frequency of the moving mirror and the fundamental
mode of the static cavity $\omega_{1}^{\mathcal{C}}\equiv\omega_{1}$. We
restrict our analysis of the effective Hamiltonian (\ref{8}) to a first order
approximation in $\varepsilon$ since we must estimate both quantities of
interest, the average number of photon creation (\ref{9}) and the linear
entropy (\ref{11}), going with $H_{I}^{2}$ to second order in $\varepsilon$.

Remembering that $\xi_{k}^{\mathcal{C}}(t),\mu_{k\ell}^{\mathcal{CS}%
}(t)\propto\dot{q}(t)$ and $\mathcal{O}_{\mathcal{S},k}(t)=a_{\mathcal{S}%
,k}\exp\left[  -i\Omega_{k}^{\mathcal{S}}(t)\right]  $, with $\Omega
_{k}^{\mathcal{S}}(t)=%
{\textstyle\int\nolimits_{0}^{t}}
d\tau$ $\omega_{k}^{\mathcal{S}}(\tau)\simeq\omega_{k}^{\mathcal{S}}%
t=k\omega_{1}^{\mathcal{S}}t$ to zeroth order in $\varepsilon$, the effective
Hamiltonian in the interaction picture (\ref{8}) can be represented as
\begin{align*}
H_{I}  &  =i\sum_{k}\left\{  \xi_{k}^{\mathcal{C}}(t)\left(  \mathcal{O}%
_{\mathcal{C},k}^{\dagger}(t)\right)  ^{2}+\sum_{\ell(\neq k)}\sum
_{\mathcal{S}}\mu_{k\ell}^{\mathcal{CS}}(t)\mathcal{O}_{\mathcal{C}%
,k}^{\dagger}(t)\left[  \mathcal{O}_{\mathcal{S},\ell}^{\dagger}%
(t)+\mathcal{O}_{\mathcal{S},\ell}(t)\right]  -H.c.\right\} \\
&  \propto\sum_{k}\dot{q}(t)\left\{  \operatorname*{e}\nolimits^{ik\omega
_{1}t}\left[  \operatorname*{e}\nolimits^{ik\omega_{1}t}\left(  a_{\mathcal{C}%
,k}^{\dagger}\right)  ^{2}+\sum_{\ell(\neq k)}\sum_{\mathcal{S}}\left(
\operatorname*{e}\nolimits^{i\ell\omega_{1}^{\mathcal{S}}t}a_{\mathcal{C}%
,k}^{\dagger}a_{\mathcal{S},\ell}^{\dagger}+\operatorname*{e}\nolimits^{-i\ell
\omega_{1}^{\mathcal{S}}t}a_{\mathcal{C},k}^{\dagger}a_{\mathcal{S},\ell
}\right)  \right]  -H.c.\right\}  \text{.}%
\end{align*}
Since $\dot{q}(t)\propto\left(  \operatorname*{e}\nolimits^{i\mathfrak{p}%
\omega_{1}t}+\operatorname*{e}\nolimits^{-i\mathfrak{p}\omega_{1}t}\right)  $,
a rotating-wave approximation gives us the leading terms%

\begin{align*}
H_{I}  &  \propto\sum_{k}\left\{  \operatorname*{e}\nolimits^{-i\left(
\mathfrak{p}-2k\right)  \omega_{1}t}\left(  a_{\mathcal{C},k}^{\dagger
}\right)  ^{2}+\sum_{\ell(\neq k)}\left[  \operatorname*{e}%
\nolimits^{-i\left[  \mathfrak{p}-\left(  k+\ell\right)  \right]  \omega_{1}%
t}a_{\mathcal{C},k}^{\dagger}a_{\mathcal{C},\ell}^{\dagger}+\operatorname*{e}%
\nolimits^{-i\left[  \mathfrak{p}-\left(  k+\ell\varkappa\right)  \right]
\omega_{1}t}a_{\mathcal{C},k}^{\dagger}a_{\mathcal{R},\ell}^{\dagger}\right.
\right. \\
&  \left.  \left.  +\left(  \operatorname*{e}\nolimits^{i\left[
\mathfrak{p}+\left(  k-\ell\right)  \right]  \omega_{1}t}+\operatorname*{e}%
\nolimits^{-i\left[  \mathfrak{p}-\left(  k-\ell\right)  \right]  \omega_{1}%
t}\right)  a_{\mathcal{C},k}^{\dagger}a_{\mathcal{C},\ell}+\left(
\operatorname*{e}\nolimits^{i\left[  \mathfrak{p}+\left(  k-\ell
\varkappa\right)  \right]  \omega_{1}t}+\operatorname*{e}\nolimits^{-i\left[
\mathfrak{p}-\left(  k-\ell\varkappa\right)  \right]  \omega_{1}t}\right)
a_{\mathcal{C},k}^{\dagger}a_{\mathcal{R},\ell}\right]  +H.c.\right\}
\end{align*}
where $\varkappa=\omega_{1}^{\mathcal{R}}/\omega_{1}^{\mathcal{C}}$. From the
above Hamiltonian, we identify five distinct resonant processes. The first one
$i)$ comes from the terms associated with the degenerate ($k=\ell$)
photon-pair creation, which contributes significantly only for even integers,
$\mathfrak{p}_{even}=2k$, feeding all modes $k=\mathfrak{p}_{even}/2$. (For
the odd integers, the occurrence of degenerate photon-pair creation requires
the nonphysical semi-integer modes $k=\mathfrak{p}_{odd}/2$.) In this case, we
have the minimum value $\mathfrak{p}_{even}^{\min}=2$. Two distinct processes
are associated with nondegenerate ($k\neq\ell$) photon-pair creation: $ii)$
the creation of both photons in\ cavity-dominated modes and $iii)$ the
creation of one photon in a cavity-dominated mode and the other in a
reservoir-dominated mode. The terms associated with the former case contribute
significantly for both even and odd integers, feeding the modes $k+\ell
=\mathfrak{p}_{even}$ or $k+\ell=\mathfrak{p}_{odd}$, respectively. Here,
since $k\neq\ell$, we obtain the minimum values $\mathfrak{p}_{even}^{\min}=4$
and $\mathfrak{p}_{odd}^{\min}=3$. The terms associated with the later case
contributes for nonintegers values of $\mathfrak{p}$, feeding the modes $k$
and $\ell$ satisfying the relation $k+\ell\varkappa$ $=$ $\mathfrak{p}$.

The two remaining processes are associated with the scattering of photons from
one cavity-dominated mode $iv)$ to another and $v)$ to a reservoir-dominated
mode. The terms associated with the former case contribute significantly for
integers $\mathfrak{p}=\left\vert k-\ell\right\vert $, scattering photons from
mode $k$ ($\ell$) to $\ell$ ($k$) if $k>\ell$ ($k<\ell$). Evidently,
$\mathfrak{p}_{even}^{\min}=2$ and $\mathfrak{p}_{odd}^{\min}=1$. The terms
associated with the later case contributes significantly for nonintegers
$\mathfrak{p}=\left\vert k-\ell\varkappa\right\vert $.

It is worth nothing that integers $\mathfrak{p}$ enable the resonant processes
$i)$, $ii)$, and $iv)$, while the nonintegers $\mathfrak{p}$ enables the
processes $iii)$ and $v)$.

\section{The average number of photon creation for the oscillatory motion of
the cavity mirror}

In this section we compute the average number of photon creation under the
particular law of motion (\ref{12}) and so, the resonances specified above.
Starting from the general Eq. (\ref{9}) and considering a thermal distribution
for the $k$th cavity-dominated mode, we obtain, to second order in
$\varepsilon$, the expression
\begin{align}
\Delta N_{\mathcal{C},k}(\tau)  &  =\left(  \frac{\mathfrak{p}\Gamma\tau}%
{4}\right)  ^{2}\left[  2N_{\mathcal{C},k}(0)+1\right]  \delta_{\mathfrak{p}%
,2k}\nonumber\\
&  +\left(  \frac{\mathfrak{p}\tau}{4}\right)  ^{2}\sum_{\ell(\neq k)}\left\{
\left[  N_{\mathcal{C},k}(0)+N_{\mathcal{C},\ell}(0)+1\right]  \left[
\mathfrak{M}_{k\ell}^{\mathcal{CC}}(0)\right]  ^{2}\frac{\left(
k-\ell\right)  ^{2}}{k\ell}\delta_{\mathfrak{p},k+\ell}\right. \nonumber\\
&  -\left[  N_{\mathcal{C},k}(0)-N_{\mathcal{C},\ell}(0)\right]  \left[
\mathfrak{M}_{k\ell}^{\mathcal{CC}}(0)\right]  ^{2}\frac{\left(
k+\ell\right)  ^{2}}{k\ell}\theta\left(  \mathfrak{p}\right)  \left(
\delta_{\mathfrak{p},k-\ell}+\delta_{\mathfrak{p},\ell-k}\right) \nonumber\\
&  -\left[  N_{\mathcal{R},k}(0)+N_{\mathcal{C},\ell}(0)+1\right]  \left[
\mathfrak{M}_{k\ell}^{\mathcal{CR}}(0)\right]  ^{2}\frac{k}{\ell\varkappa
}\delta_{\mathfrak{p},k+\ell\varkappa}\nonumber\\
&  \left.  +\left[  N_{\mathcal{C},k}(0)-N_{\mathcal{R},\ell}(0)\right]
\left[  \mathfrak{M}_{k\ell}^{\mathcal{CR}}(0)\right]  ^{2}\frac{k}%
{\ell\varkappa}\theta\left(  \mathfrak{p}\right)  \left(  \delta
_{\mathfrak{p},k-\ell\varkappa}+\delta_{\mathfrak{p},\ell\varkappa-k}\right)
\right\}  \text{,} \label{N}%
\end{align}
where we have defined the dimensionless time variable $\tau=\varepsilon
\omega_{1}t$, the effective coupling strengths%
\begin{align*}
\Gamma &  =1-\frac{1}{\pi}\eta_{1}+O(\eta_{1}^{2})\text{,}\\
\mathfrak{M}_{k\ell}^{\mathcal{CS}}(t)  &  =q_{0}\int_{-L_{0}}^{q(t)}%
dx\frac{\partial\psi_{k}^{\mathcal{C}}(x,t)}{\partial q}\psi_{\ell
}^{\mathcal{S}}(x,t)\text{,}%
\end{align*}
apart from the step function%
\[
\theta\left(  x\right)  =\left\{
\begin{tabular}
[c]{ll}%
$0$ & $x\leq0$\\
$1$ & $x>0$%
\end{tabular}
\ \ \ \ \right.  \text{.}%
\]
\qquad

From the expression derived above for the average number of photon creation,
which generalizes previous development in literature, we first observe that
for an static cavity where $\varepsilon=0$ or $\mathfrak{p}=0$, we end up with
the expected result $\Delta N_{\mathcal{C},k}(\tau)=0$. Moreover, the five
distinct resonant processes identified above becomes evident: the first three
terms on the right hand side of Eq. (\ref{N}) correspond to the cases $i)$,
$ii)$, and $iv)$, whereas the remaining two terms correspond to the cases
$iii)$ and $v)$, respectively. Interestingly, we observe that the scattering
processes only take place, up to second order in $\varepsilon$, when
temperature effects are taking into account.

\subsection{The parametric amplification process}

For the particular case $\mathfrak{p}=2$, we obtain for the fundamental mode
$k=1$, the result%

\begin{equation}
\Delta N_{\mathcal{C},1}(\tau)=\left(  \frac{\Gamma\tau}{2}\right)
^{2}\left[  2N_{\mathcal{C},1}(0)+1\right]  -\frac{4}{3}\tau^{2}\left(
N_{\mathcal{C},1}(0)-N_{\mathcal{C},3}(0)\right)  \left[  \mathfrak{M}%
_{1,3}^{\mathcal{CC}}(0)\right]  ^{2}\text{,} \label{N2}%
\end{equation}
showing the degenerate photon-pair creation apart from photon scattering from
the fundamental mode to the third one, as dictated by processes $i)$ and
$iv)$, respectively. We observe that the nondegenerate photon-pair creation
$ii)$ does not occur since the condition $\mathfrak{p}=k+\ell$ is not
satisfied. As expected, the number $\Delta N_{\mathcal{C},1}(\tau)$ grows
linearly with the temperature. For the case where the \textquotedblleft cavity
$+$ reservoir\textquotedblright\ system is at absolute zero the result in Eq.
(\ref{N2}), simplifies to
\[
\Delta N_{\mathcal{C},1}(\tau)=\left(  \frac{\Gamma\tau}{2}\right)
^{2}\text{,}%
\]
which recovers the result in Ref. \cite{Dodo} for an ideal cavity where
$\Gamma=1$.

\subsection{The case $\mathfrak{p}=1$}

We finally note that, although we do not have photon creation for the case
$\mathfrak{p}=1$, where
\begin{equation}
\Delta N_{\mathcal{C},1}(\tau)=-\frac{9}{32}\tau^{2}\left[  N_{\mathcal{C}%
,1}(0)-N_{\mathcal{C},2}(0)\right]  \left[  \mathfrak{M}_{1,2}^{\mathcal{CC}%
}(0)\right]  ^{2}\text{,} \label{N3}%
\end{equation}
we do have photon scattering from the fundamental mode to the second one,
through the process $iv)$, when $N_{\mathcal{C},1}(0)=\operatorname{Tr}\left(
\rho_{\mathcal{C},1}(0)a_{\mathcal{C},1}^{\dagger}a_{\mathcal{C},1}\right)
\neq0$, i.e., when there is initial excitation in the fundamental mode.

\subsection{Graphical Analysis}

We next present the graphical results for the average number of
photons\ created $\Delta N_{\mathcal{C},k}(\tau)$, computed through Eq.
(\ref{N}). To this end we consider integers $\mathfrak{p}$, an absolute zero
reservoir, and all the cavity-dominated modes also in the vacuum state. We
start by plotting, in Fig. $2$(a) and (b), $\Delta N_{\mathcal{C},k}(\tau)$
versus $k$ for fixed values $\mathfrak{p}=14$ and $15$, respectively. As we
conclude from the phenomenological analysis of the effective Hamiltonian
(\ref{8}) leading to Eq. (\ref{N}), the modes which are fed by photon creation
are those where $k\leq\mathfrak{p}-1$, while the maximum number of photon
creation occurs in the single mode $k=7$ for $\mathfrak{p}=14$ and in the pair
of modes $\left(  k,\ell\right)  =\left(  6,7\right)  $ for $\mathfrak{p}=15$.
From Fig. $2$ to $4$ we fixed the time interval $\tau=1/\mathfrak{p}$, during
which the mirror performs $1/2\pi\varepsilon$ oscillations for all values of
$\mathfrak{p}$.

Instead of fixing $\mathfrak{p}$, in Fig. 3 we fixed $k=7$ to plot $\Delta
N_{\mathcal{C},7}(\tau)$ versus $\mathfrak{p}$. We verify, as expected, the
occurrence of resonances in the average number of photon creation for
$\mathfrak{p}\geq8$, with a maximum $\left\langle N_{\mathcal{C},7}%
(\tau)\right\rangle _{\max}$ for $\mathfrak{p}=14$. As evidenced in Fig. 3,
with the exception of the maximum for $\mathfrak{p}=2k$, the magnitude of
$\Delta N_{\mathcal{C},7}(\tau)$ exhibits a profile governed by the effective
coupling matrix elements $\mathfrak{M}_{k\ell}^{\mathcal{CC}}(0)$ apart from
the ratios $\left(  k\pm\ell\right)  ^{2}/k\ell$.

In Fig. 4(a) we plot\ the total number of photons created $N_{\mathcal{C}%
}=\sum_{k}\Delta N_{\mathcal{C},k}(\tau)$, versus $\mathfrak{p}$. It is
evident from Fig. 4(a) that this number is significantly larger for the even
values of $\mathfrak{p}$, where the degenerate photon-pair creation takes
place. Moreover, as expected from our phenomenological analysis, there is no
photon creation for $\mathfrak{p}=1$ and the total number $N_{\mathcal{C}}$
increases as $\mathfrak{p}$ increases. A behavior similar to that in Fig. 4(a)
follows from the plot, in Fig. 4(b), of the total normalized energy
$E_{\mathcal{C}}=\sum_{k}\omega_{k}\Delta N_{\mathcal{C},k}(\tau)/\omega_{1}$
versus $\mathfrak{p}$. Whereas in Fig. 4(a) the number of photon creation
increases as the detuning $\mathfrak{p}$ increases, in Fig. 4(b), this
behavior is modulated by the multiplicative frequency $\omega_{k}$.

Finally, in Fig. 5 we plot $\Delta N_{\mathcal{C},1}(\tau)$ versus $\tau$ for
$\mathfrak{p}=2$ and the cases where the whole system (the cavity plus
reservoir) is at 0K, i.e., $\left\langle N_{\mathcal{S},k}(0)\right\rangle =0$
(solid line) and at $\omega_{1}/k_{B}\ln(1.1)$ K where $\left\langle
N_{\mathcal{C},1}(0)\right\rangle =10$ (dashed line). We observe that the
number of photon creation increases with the temperature, as also verified in
Ref. \cite{Soff2}, corroborating the result demonstrated by L. Parker
\cite{Parker} that the initial presence of bosons tends to increase the number
of bosons created inside the cavity

\section{Entropy and decoherence time under an oscillatory motion of the
cavity mirror}

In this section we compute, under the particular law of motion (\ref{12}) and
the corresponding resonances, the entropy (\ref{11}) and the decoherence time
of a "Schr\"{o}dinger-cat"-like state $\left\vert \psi_{\mathcal{C}%
,k}(0)\right\rangle =\mathcal{N}\left(  \left\vert \alpha_{0}\right\rangle
+\left\vert -\alpha_{0}\right\rangle \right)  $ prepared in the $k$th
cavity-dominated mode of an initially static cavity. To second order in
$\varepsilon$, we obtain for the entropy%
\begin{align}
S_{\mathcal{C},k}(\tau)  &  \simeq\left(  \mathfrak{p}\tau\right)  ^{2}%
\sum_{\ell(\neq k)}\left\{  \left(  \left\vert \alpha\right\vert
^{2}+1\right)  \left[  \mathfrak{M}_{k\ell}^{\mathcal{CC}}(0)\right]
^{2}\frac{\left(  k-\ell\right)  ^{2}}{2k\ell}\delta_{p,\ell+k}\right.
\nonumber\\
&  +\left\vert \alpha\right\vert ^{2}\left[  \mathfrak{M}_{k\ell
}^{\mathcal{CC}}(0)\right]  ^{2}\frac{\left(  k+\ell\right)  ^{2}}{2k\ell
}\left(  \delta_{p,k-\ell}+\delta_{p,\ell-k}\right)  \theta\left(  p\right)
\nonumber\\
&  -\left(  \left\vert \alpha\right\vert ^{2}+1\right)  \frac{k}%
{2\ell\varkappa}\left[  \mathfrak{M}_{k\ell}^{\mathcal{CR}}(0)\right]
^{2}\delta_{p,k+\ell\varkappa}\nonumber\\
&  \left.  -\left\vert \alpha\right\vert ^{2}\frac{k}{2\ell\varkappa}\left[
\mathfrak{M}_{k\ell}^{\mathcal{CR}}(0)\right]  ^{2}\left(  \delta
_{p,k-\ell\varkappa}+\delta_{p,\ell\varkappa-k}\right)  \theta\left(
p\right)  \right\}  \text{,} \label{S}%
\end{align}
which can be considered to estimate the decoherence time through the relation%
\begin{equation}
S_{\mathcal{C},k}^{\mathfrak{p}}(\tau)\simeq\left(  \frac{\tau}{\tau_{D}%
}\right)  ^{2}\text{,} \label{TD}%
\end{equation}
connected to the \textquotedblleft idempotency defect\textquotedblright\ in
Ref. \cite{Piza}.

We first observe that, for the particular case where the $k$th
cavity-dominated mode is also in the vacuum state, as all other modes, we
still have an increase of the entropy, given by%
\begin{align*}
S_{\mathcal{C},k}(\tau)  &  \simeq\left(  \frac{\mathfrak{p}\tau}{2}\right)
^{2}\sum_{\ell(\neq k)}\left\{  \left[  \mathfrak{M}_{k\ell}^{\mathcal{CC}%
}(0)\right]  ^{2}\frac{\left(  k-\ell\right)  ^{2}}{2k\ell}\delta_{p,\ell
+k}\right. \\
&  \left.  -\frac{k}{2\ell\varkappa}\left[  \mathfrak{M}_{k\ell}%
^{\mathcal{CR}}(0)\right]  ^{2}\delta_{p,k+\ell\varkappa}\right\}  \text{.}%
\end{align*}
Evidently, in this case, the increase of the entropy comes entirely from
photon creation.

Now, considering the initial state $\left\vert \psi_{\mathcal{C}%
,1}(0)\right\rangle =\mathcal{N}\left(  \left\vert \alpha_{0}\right\rangle
+\left\vert -\alpha_{0}\right\rangle \right)  $, prepared in the fundamental
mode, we obtain for $\mathfrak{p}=1$ and $\mathfrak{p}=2$, the results
\begin{align*}
S_{\mathcal{C},1}^{\mathfrak{p}=1}(\tau)  &  \simeq\frac{9}{4}\tau
^{2}\left\vert \alpha\right\vert ^{2}\left[  \mathfrak{M}_{1,2}^{\mathcal{CC}%
}(0)\right]  ^{2}\text{,}\\
S_{\mathcal{C},1}^{\mathfrak{p}=2}(\tau)  &  \simeq\frac{32}{3}\tau
^{2}\left\vert \alpha\right\vert ^{2}\left[  \mathfrak{M}_{1,3}^{\mathcal{CC}%
}(0)\right]  ^{2}\text{,}%
\end{align*}
associated with the decoherence times
\begin{align*}
\tau_{D}^{\mathfrak{p}=1}  &  \simeq\frac{2}{3}\frac{1}{\left\vert
\alpha\right\vert \mathfrak{M}_{1,2}^{\mathcal{CC}}(0)}\text{,}\\
\tau_{D}^{\mathfrak{p}=2}  &  \simeq\sqrt{\frac{3}{32}}\frac{1}{\left\vert
\alpha\right\vert \mathfrak{M}_{1,3}^{\mathcal{CC}}(0)}\text{.}%
\end{align*}
We note that both decoherence time $\tau_{D}^{\mathfrak{p}=1}$ and $\tau
_{D}^{\mathfrak{p}=2}$ are entirely due to photon scattering from the
fundamental to the first and third excited modes, respectively, governed by
the process $iv)$. Therefore, as expected, we get a larger-than-unity ratio
\[
\frac{\tau_{D}^{\mathfrak{p}=1}}{\tau_{D}^{\mathfrak{p}=2}}\simeq\sqrt
{\frac{3}{2}}\left(  1-\frac{1}{\pi}\eta_{1}\right)  +O(\eta_{1}^{2})\text{.}%
\]
which recovers, for the particular case of an ideal cavity ($\gamma
\rightarrow\infty$),\ exactly the result obtained in Ref. \cite{Dodonov1},
through a different technique.

Finally, we verify that the decoherence times for the ideal and nonideal DCEs
satisfy the ratio%
\[
\frac{\tau_{D}^{\mathfrak{p}}(\gamma)}{\tau_{D}^{\mathfrak{p}}(\gamma
\rightarrow\infty)}\simeq1-\frac{1}{\pi}\eta_{1}+O(\eta_{1}^{2})\text{,}%
\]
for the particular cases $\mathfrak{p}=1$ and $\mathfrak{p}=2$.

For a graphical analysis of the entropy and, consequently, the loss of purity
and decoherence in the nonideal DCE, we assume the\ "Schr\"{o}dinger-cat"-like
state $\left\vert \psi_{\mathcal{C},1}(0)\right\rangle =\mathcal{N}\left(
\left\vert \alpha_{0}\right\rangle +\left\vert -\alpha_{0}\right\rangle
\right)  $ to be prepared in the fundamental mode. As we conclude from Fig. 6,
where the entropy $S_{\mathcal{C},1}(\tau)$ is plotted against $\mathfrak{p}$
for $\left\vert \alpha_{0}\right\vert ^{2}=$ $2$ and the time intervals
$\tau=0.1$ (black triangles) and $\tau=0.25$ (black circles), the purity loss
of the initial state $\left\vert \psi_{\mathcal{C},1}(0)\right\rangle $
exhibits resonances, as does the average number of photon creation. The
increase of the entropy with $\mathfrak{p}$, follows from the process of
nondegenerate photon-pair creation, occurring for $\mathfrak{p}\geq3$, and
photon scattering, occurring for $\mathfrak{p}\geq1$. In fact, these processes
couple the fundamental mode where the "Schr\"{o}dinger-cat"-like state is
prepared ($k=1$) to other cavity-dominated modes ($\ell=\mathfrak{p}\pm1\geq
2$), thus increasing its entropy when tracing out the remaining $\ell$ modes.
Moreover, as we observe from Eqs. (\ref{S}) and (\ref{N}), the rate of both
processes\ of nondegenerate photon-pair creation,and photon scattering
increases with $\mathfrak{p}$. We also observe that the degenerate photon-pair
creation process in $k=1$, occurring when $\mathfrak{p}=2$, does not couple
$k=1$ to any other cavity-dominated mode and, consequently, does not increases
its entropy. We finally note that the inclination which characterizes the
linear increase of the entropy with $\mathfrak{p}$ decreases when $\gamma$
increases, as seen in Fig. 6 from the grey triangles and grey circles
associated respectively with $\tau=0.1$ and $\tau=0.25$. For an ideal cavity,
the entropy $S_{\mathcal{C},1}(\tau)$ reduces to the expression%
\begin{equation}
S_{\mathcal{C},1}(\tau)\simeq\frac{\tau^{2}}{2}\times\left\{
\begin{array}
[c]{ccc}%
2\left\vert \alpha\right\vert ^{2}\mathfrak{p}+\mathfrak{p}-1 & \text{for} &
\mathfrak{p}\geq3\\
\left\vert \alpha\right\vert ^{2}\left(  \mathfrak{p}+1\right)  & \text{for} &
\mathfrak{p}=1,2
\end{array}
\right.  \text{,}%
\end{equation}
for integers $\mathfrak{p}$.

Next, in Fig. 7, we plot the normalized decoherence time $\tau_{D}%
^{\mathfrak{p}}/\tau_{D}^{\mathfrak{p}=1}$ of the "Schr\"{o}dinger-cat"-like
state $\left\vert \psi_{\mathcal{C},1}(0)\right\rangle $ against
$\mathfrak{p}$. As expected from Fig. 6, the decoherence time of the state
$\left\vert \psi_{\mathcal{C},1}(0)\right\rangle $ decreases as $\mathfrak{p}$
increases. The resonances in the nonideal DCE indicate that there is
practically no purity loss and decoherence when the frequency of the moving
mirror does not fit the resonance conditions specified above. In fact, for an
off-resonance DCE, the effective coupling between the cavity modes and the
number of photon creation and scattering are practically null, protecting the
prepared state, whatever its selected mode. Therefore, for an off-resonance
DCE, the prepared superposition $\left\vert \psi_{\mathcal{C},1}%
(0)\right\rangle $ becomes a nonstationary state, following the dynamics
governed by the moving mirror, but protected from the decoherence mechanisms
present in the on-resonant regime.

\section{Concluding remarks}

Considering the problem of the nonideal DCE, we constructed a dissipative
effective Hamiltonian, Eq. (\ref{8}), which applies to any law of motion for
the boundary. From this Hamiltonian, we compute a general expression, in Eq.
(\ref{9}), for the average number of photon creation in the $k$th
cavity-dominated mode, $\Delta N_{\mathcal{C},k}(\tau)$, which also applies
for any law of motion of the cavity mirror. When considering a particular
oscillatory law of motion of the mirror, as in Eq. (\ref{12}), the expression
(\ref{N}) enables us to present a comprehensive analysis of the resonances in
the number of photon creation --- as demonstrated by Figs. $2$, $3$, and $4$
--- since it applies to any value of the detuning $\mathfrak{p}$ between the
frequency of the moving mirror and the fundamental mode. In the literature,
even when addressing the ideal DCE, the authors confine themselves to specific
law of motion and proceed to the analysis of photon creation only within
particular values of $\mathfrak{p}$. The generalized expression we have
derived in Eq. (\ref{9}) for the average number of photon creation can
certainly be useful for further investigations of the DCE.

We also present a general treatment of the linear entropy of the evolved $k$th
cavity-dominated state, $S_{\mathcal{C},k}(t)=1-\operatorname*{Tr}%
\rho_{\mathcal{C},k}^{2}(t)$, given by Eq. (\ref{11}), also applicable for any
law of motion of the mirror. Considering again the particular law of motion
(\ref{12}), we have computed through the linear entropy, the decoherence time
of a\ "Schr\"{o}dinger-cat"-like superposition of coherent states initially
prepared in the $k$th cavity-dominated mode of the static cavity. To this end,
we have used the expression (\ref{TD}) to estimate the decoherence time
through a second-order expansion of the entropy, presenting a comprehensive
analysis of the decoherence process within the nonideal DCE and retrieving
previous particular results reported in the literature \cite{Dodonov1}.

We expect the present approach to be useful for further investigations of the
nonideal DCE, considering for example a more realistic reservoir acting even
on a static cavity. In fact, the derivation of a more realistic reservoir for
the nonideal DCE is by itself an interesting task.

We wish to express thanks for the support from the Brazilian agencies FAPESP
and CNPq.

Fig. 1 Schematic sketch of the problem, with the nonideal cavity in the region
between $x=0$ and the moving mirror $x=q(t)$, and the reservoir in the region
where $-L_{0}\leq x\leq0$.

Fig. 2 The average number of photon creation $\Delta N_{\mathcal{C},k}(\tau)$
versus $k$ for fixed values (a) $\mathfrak{p}=14$ and (b) $\mathfrak{p}=15$
and fixed time interval $\tau=1/\mathfrak{p}$.

Fig. 3 The average number of photon creation $\Delta N_{\mathcal{C},k}(\tau)$
versus $\mathfrak{p}$ for fixed $k=7$ and time interval $\tau=1/\mathfrak{p}$.

Fig. 4 (a) the total number of photons created $N_{\mathcal{C}}(\tau)$ and (b)
the total normalized energy $E_{1}(\tau)/\omega_{1}$ versus $\mathfrak{p}$,
fixed time interval $\tau=1/\mathfrak{p}$.

Fig. 5 The average number of photon creation $\Delta N_{\mathcal{C},1}(\tau)$
versus $\tau$ for $\mathfrak{p}=2$.and the cases where the whole system is at
0K (solid line) and $\omega_{1}/k_{B}\ln(1.1)$ K (dashed line), for fixed time
interval $\tau=1/\mathfrak{p}$.

Fig. 6 The linear entropy $S_{\mathcal{C},k}(\tau)$ of the
"Schr\"{o}dinger-cat"-like state $\left\vert \psi_{\mathcal{C},1}%
(0)\right\rangle $, with $\left\vert \alpha_{0}\right\vert ^{2}=$ $2$, against
$\mathfrak{p}$, for fixed time interval $\tau=0.25$.

Fig. 7 The normalized decoherence time $\tau_{D}^{\mathfrak{p}}/\tau
_{D}^{\mathfrak{p}=1}$ of the "Schr\"{o}dinger-cat"-like state $\left\vert
\psi_{\mathcal{C},1}(0)\right\rangle $ against $\mathfrak{p}$.


\begin{thebibliography}{99}                                                                                               %
\textbf{Acknowledgments}
\end{thebibliography}

\begin{thebibliography}{9999}                                                                                             %


\bibitem {Moore}G. T. Moore, J. Math. Phys. \textbf{11}, 2679 (1970).

\bibitem {Dodonov}V. V. Dodonov, A. B. Klimov, and D. E. Nikonov,
\textit{Phys. Rev. A} \textbf{47}, 4422 (1993); V. V. Dodonov, A. B. Klimov,
and V. I. Man'ko, J. Sov. Laser Res. \textbf{12}, 439 (1991).

\bibitem {4}B. S. Dewitt, Phys. Rep. \textbf{19}, 295 (1975).

\bibitem {5}S. A. Fulling and P. C. W. Davies, Proc. R. Soc. London Ser. A
\textbf{348}, 393 (1976); P. C. W. Davies and S. A. Fulling, \textit{ibid}.
\textbf{356}, 237 (1977).

\bibitem {6}E. Yablonovitch, Phys. Rev. Lett. \textbf{62}, 1742 (1989).

\bibitem {7}V. V. Hizhnyakov, Quantum Opt. \textbf{4}, 277 (1992).

\bibitem {8}S. Sarkar, Quantum Opt. \textbf{4}, 345 (1992).

\bibitem {9}V. V. Dodonov, A. B. Klimov, and D. E. Nikonov, Phys. Lett. A
\textbf{149}, 225 (1990); M. T. Jaekel and S. Reynaud, J. Phys. I (France)
\textbf{2}, 149 (1992).

\bibitem {Eberlein}C. Eberlein, Phys. Rev. Lett. \textbf{76}, 3842 (1996).

\bibitem {Knight}P. Knight, Nature \textbf{381}, 736 (1996).

\bibitem {Soff1}G. Schaller, \textit{et al}., Phys. Rev. A \textbf{66}, 023812
(2002); \textit{ibid}, Phys. Lett. A \textbf{297}, 81 (2002).

\bibitem {Soff2}G. Plunien, R. Sch\"{u}tzhold, and G. Soff, Phys. Rev. Lett.
\textbf{84}, 1882 (2000).

\bibitem {Scully}R. Lang, M. O. Scully and W. E. Lamb, Phys. Rev. A
\textbf{7}, 1788 (1973); J. Gea-Banachloche et al., Phys. Rev. A \textbf{41},
369 (1990).

\bibitem {Parker}L. Parker, Phys. Rev. Lett. \textbf{21}, 562 (1968); L.
Parker, Phys. Rev. \textbf{183}, 1057 (1969).

\bibitem {Carlitz}R. D. Carlitz and R. S. Willey, Phys. Rev. D \textbf{36},
2327 (1987).

\bibitem {Davies}P. C. W. Davies, J. Opt. B: Quantum Semiclass. Opt.
\textbf{7}, S40 (2005).

\bibitem {Fabio}F. Pascoal and C. Farina, Int. J. Theoretical Phys.
\textbf{46}, 2950 (2007).

\bibitem {Law}C. K. Law, Phys. Rev. A \textbf{49}, 433 (1994); C. K. Law,
Phys. Rev. A \textbf{51}, 2537 (1995).

\bibitem {Dodonov1}V. V. Dodonov, M. A. Andreata, and S. S. Mizrahi, J. Opt.
B: Quantum Semiclass. Opt. \textbf{7}, S468 (2005).

\bibitem {Maia}D. A. R. Dalvit and P. A. Maia Neto, Phys. Rev. Lett.
\textbf{84}, 798 (2000).

\bibitem {Zurek}W. H. Zurek, Phys. Rev. D \textbf{24}, 1516 (1981);
\textit{ibid. }\textbf{26}, 1862 (1982).

\bibitem {CL}A. O. Caldeira and A. J. Leggett, Physica \textbf{121A}, 587
(1993), \textit{ibid. }Ann. Phys. (N.Y.) 149, \textbf{374} (1983),
\textit{ibid. }Phys. Rev. A \textbf{31}, 1059 (1985).

\bibitem {QECC}P.W. Shor, Phys. Rev. A \textbf{52}, R2493 (1995); A. M.
Steane, Phys. Rev. Lett. \textbf{77}, 793 (1996).

\bibitem {PKM}J. F. Poyatos, \textit{et al}., Phys. Rev. Lett. \textbf{77},
4728 (1996); A. R. R. Carvalho, \textit{et al}., \textit{ibid.} \textbf{86},
4988 (2001); C. J. Myatt, \textit{et al}., Nature \textbf{403}, 269 (2000).

\bibitem {Lidar}D. A. Lidar and K. B. Whaley, quant-ph/0301032.

\bibitem {Mickel}M. A. de Ponte, M. C. de Oliveira, and M. H. Y. Moussa, Ann.
Phys. (N.Y.) \textbf{317}, 72 (2004); \textit{ibid}, Phys. Rev. A \textbf{70},
022324 (2004); \textit{ibid}, Phys. Rev. A \textbf{70}, 022325 (2004); M. A.
de Ponte, S. S. Mizrahi, and M. H. Y. Moussa, Ann. Phys. (N.Y.) \textbf{322},
2077 (2007); \textit{ibid}, Phys. Rev. A \textbf{76,} 032101 (2007).

\bibitem {Haroche}M. Brune, \textit{et al.}, Phys. Rev. Lett. \textbf{77},
4887 (1996).

\bibitem {Wineland}C. J. Myatt, \textit{et al}, Nature \textbf{403}, 269 (2000).

\bibitem {Walls}D. F. Walls and G. J. Milburn, \textit{Phys. Rev. A}
\textbf{31}, 2403 (1985).

\bibitem {MMC}M. H. Y. Moussa, \textit{et al}., Phys. Lett. A \textbf{221},
145 (1996).

\bibitem {Landauer}R. Landauer, Proc. R. Soc. London, Ser. A, \textbf{353},
367 (1995).

\bibitem {Unruh}W. G. Unruh, Phys. Rev. A \textbf{51}, 992 (1995).

\bibitem {DKN}V. V. Dodonov, A. B. Klimov and D. E. Nikonov, J. Math. Phys.
\textbf{34}, 2742 (1993); V. V. Dodonov, Phys. Lett. A \textbf{207}, 126 (1995).

\bibitem {Dodo}V. V. Dodonov and A. B. Klimov, Phys. Rev. A \textbf{53}, 2664 (1996).

\bibitem {Piza}J. I. Kim \textit{et al.}, Phys. Rev. Lett. \textbf{77}, 207 (1996).

\textbf{Figure captions}
\end{thebibliography}
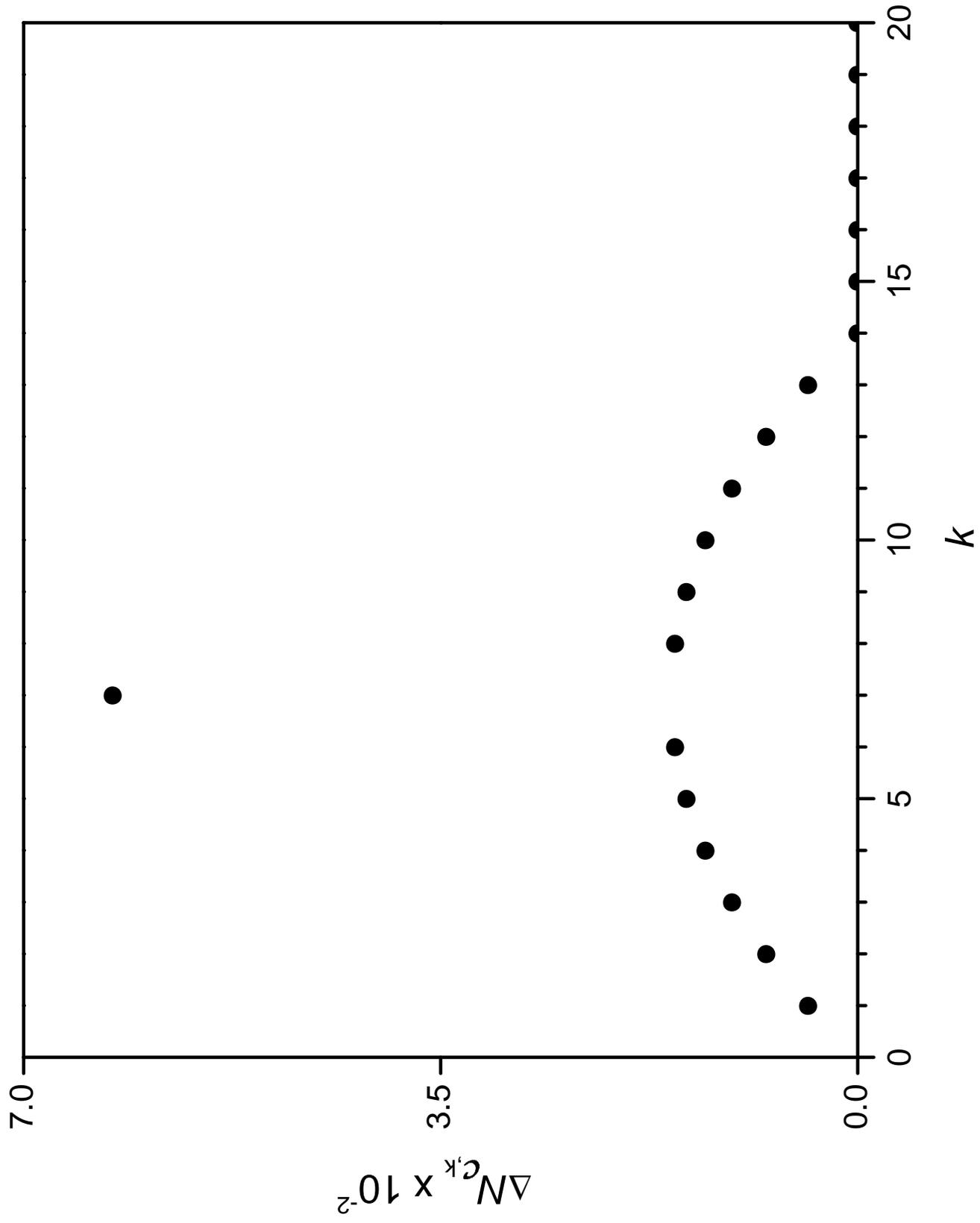
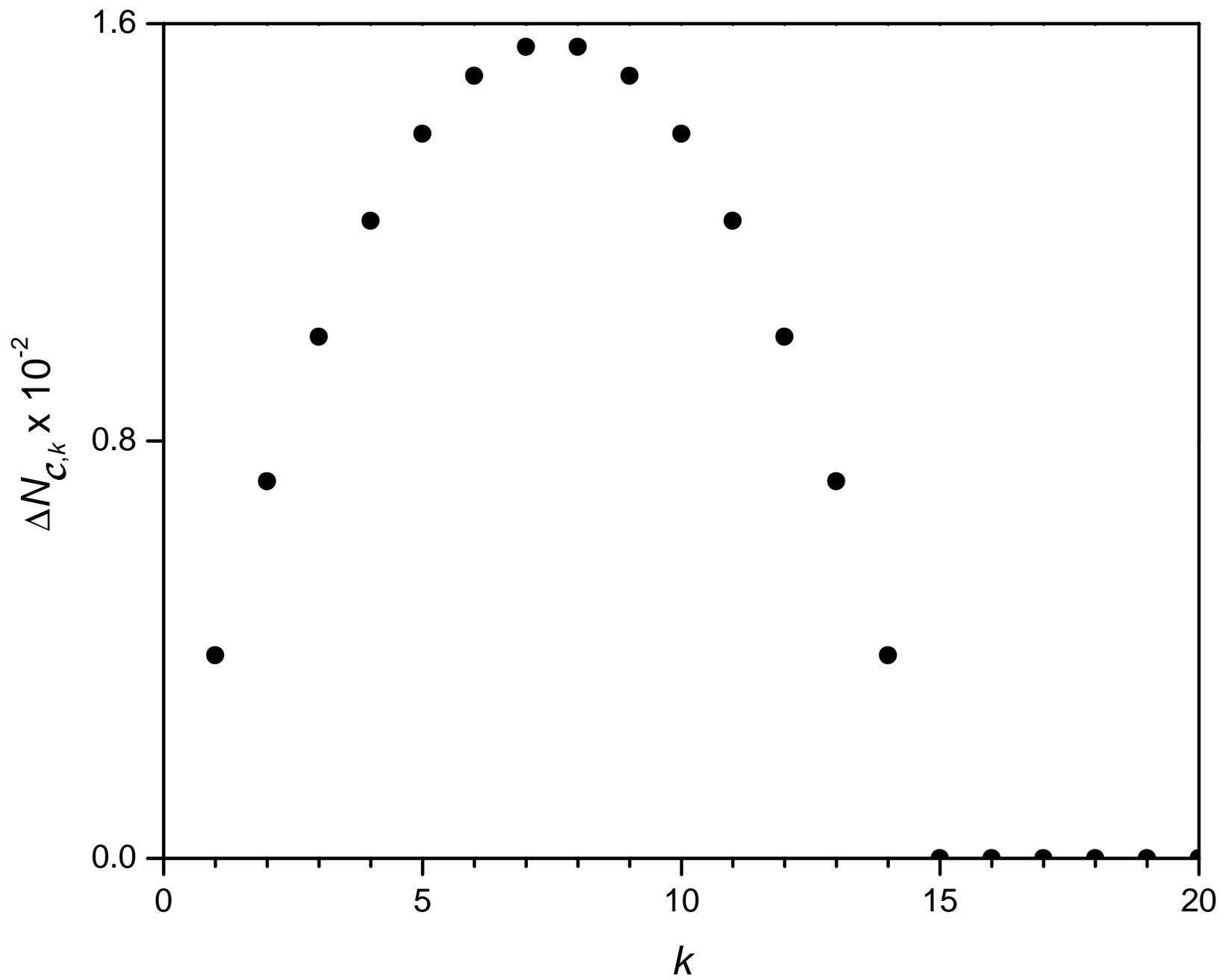
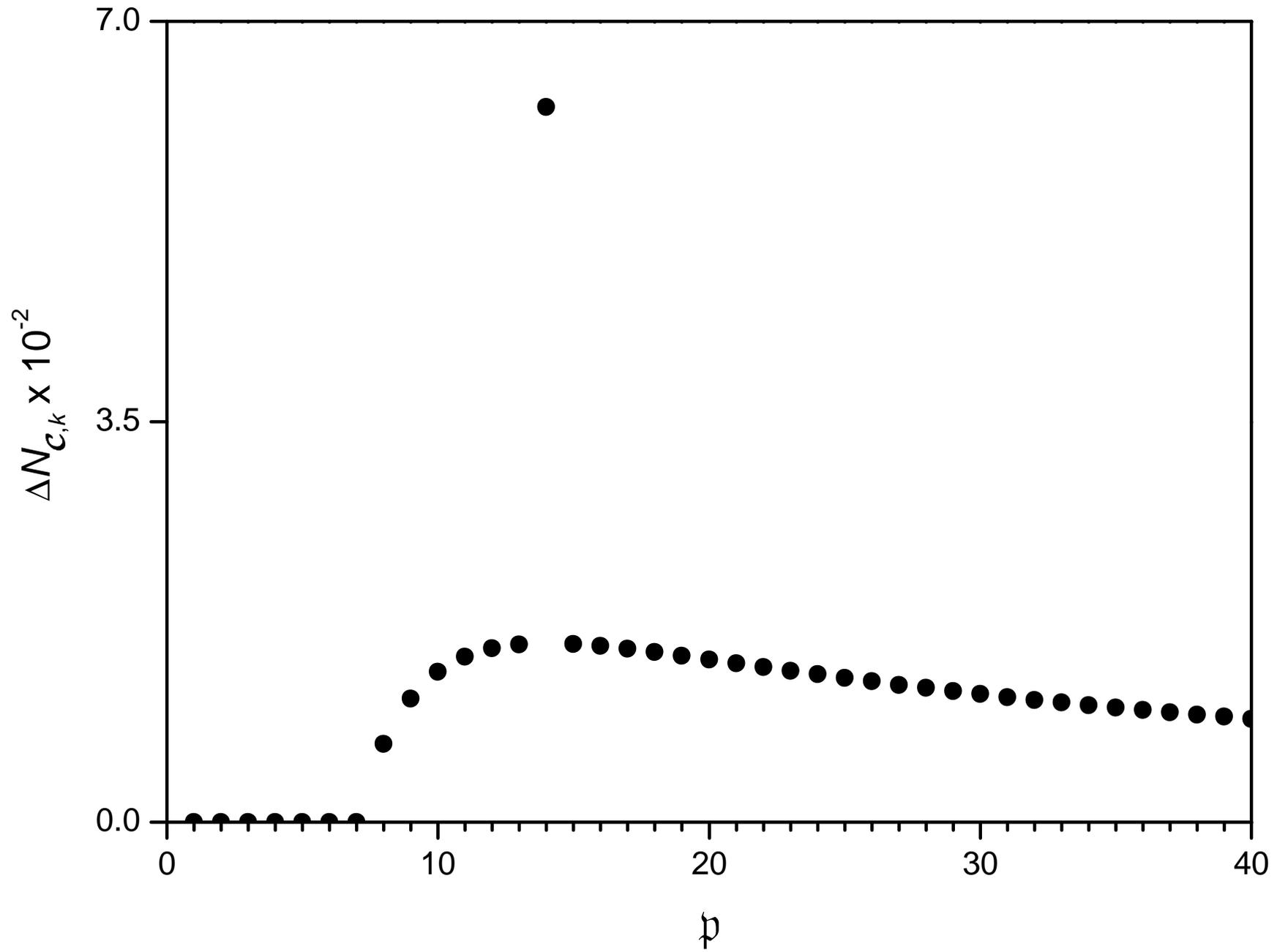
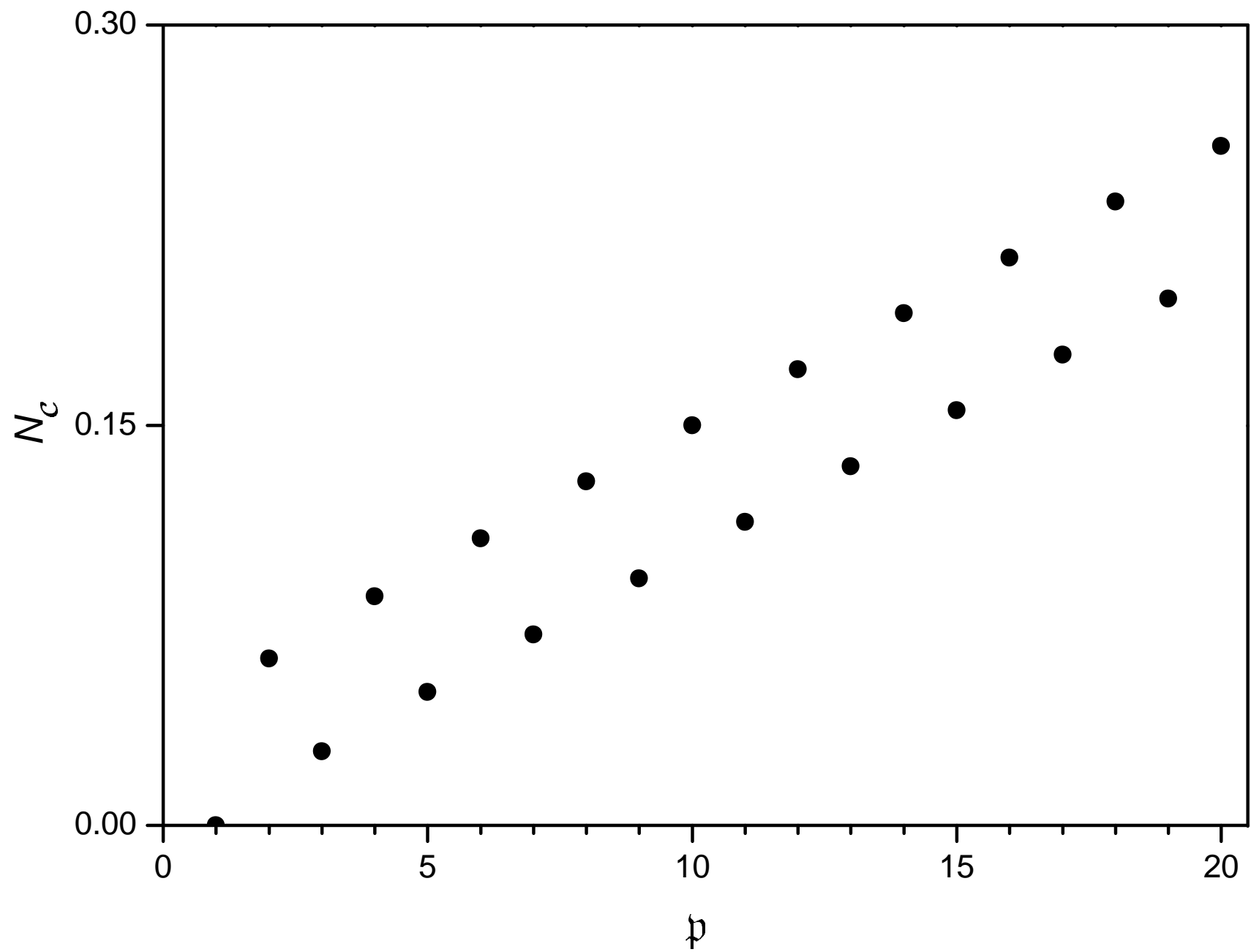
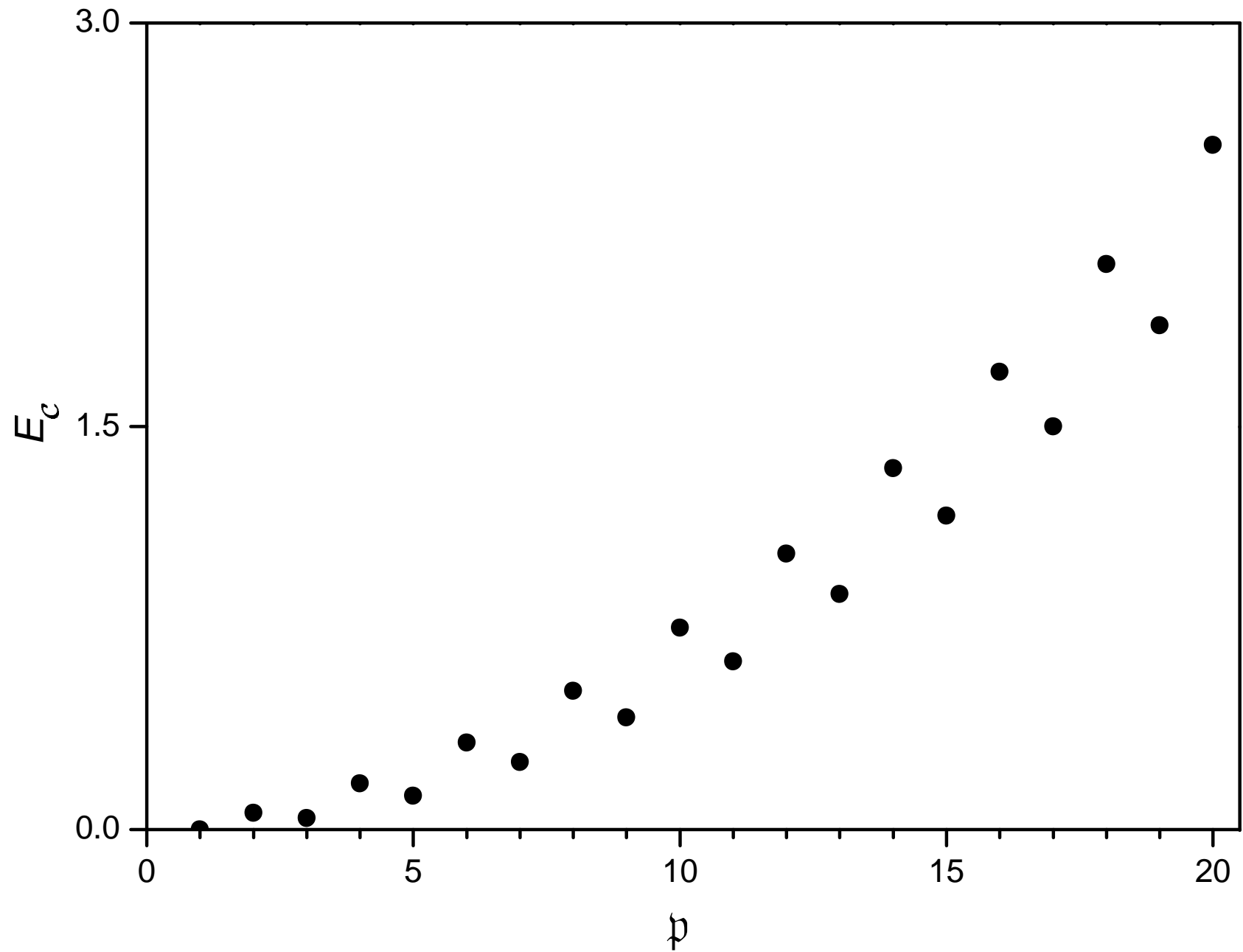
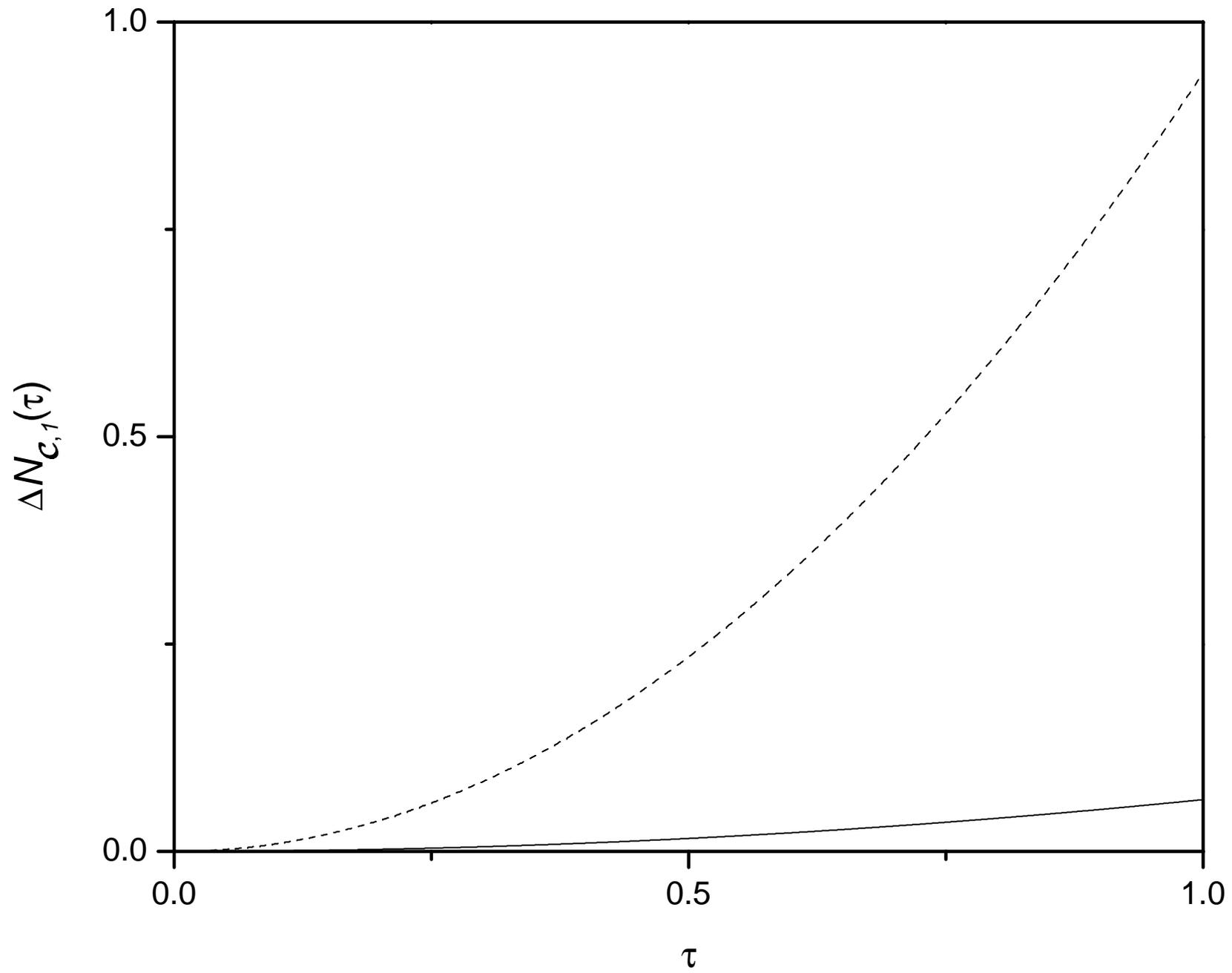
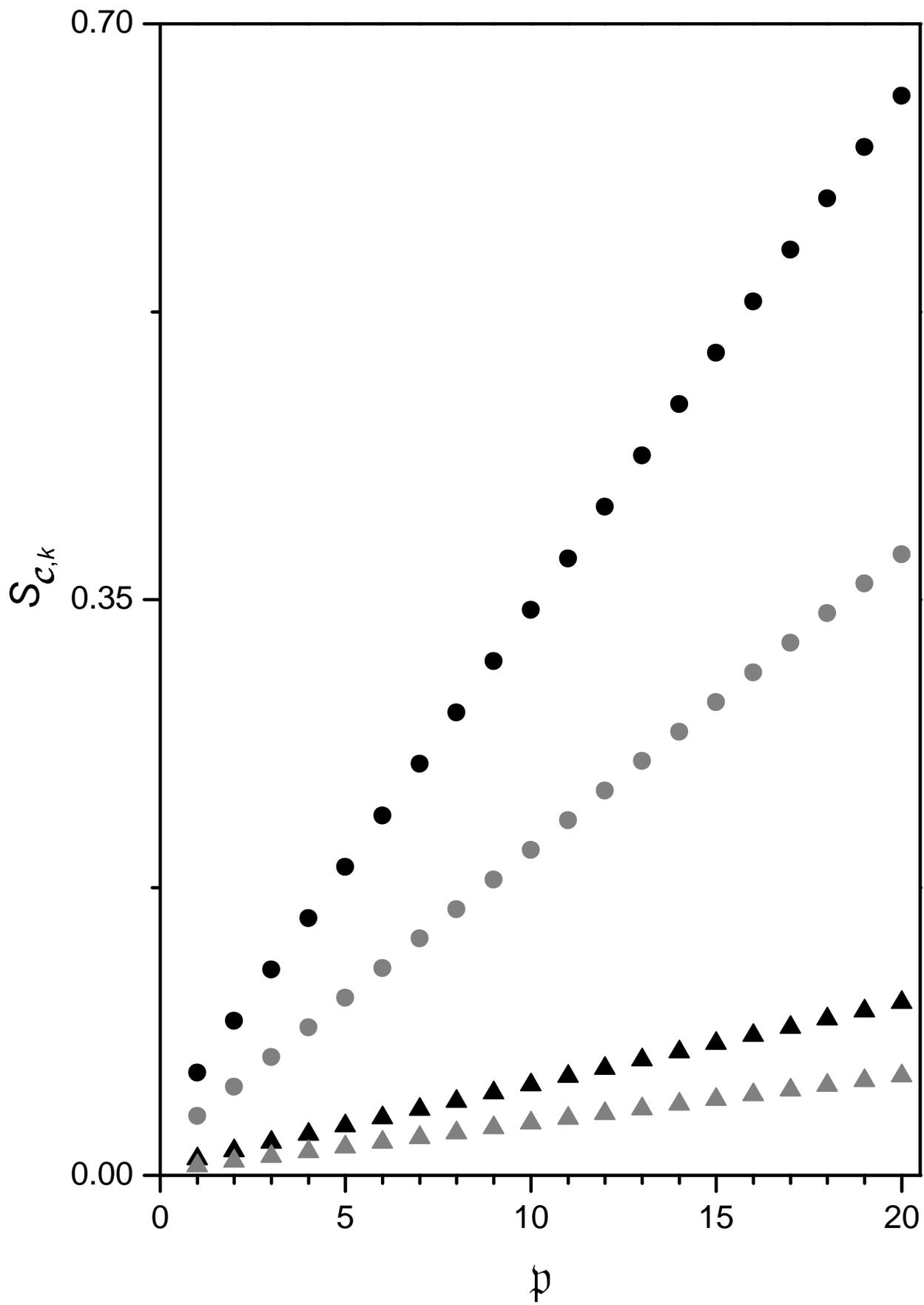
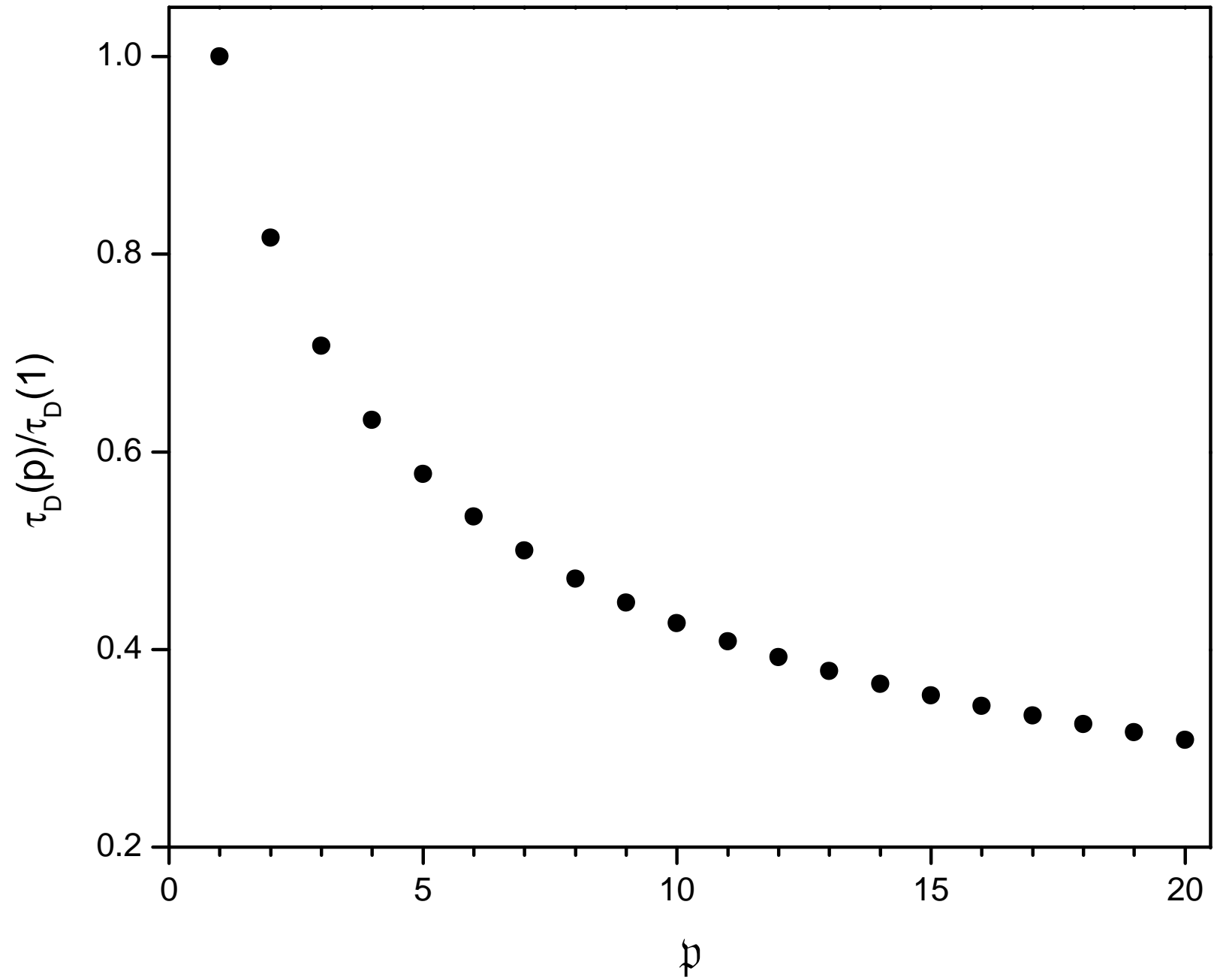
\end{document}